\documentclass{pasj00}
\usepackage[dvips]{color}
\usepackage{dcolumn}

\begin{document}
\SetRunningHead{Y. Ita et al.}{AKARI IRC survey of the LMC}
\Received{yyyy/mm/dd}
\Accepted{yyyy/mm/dd}

\title{AKARI IRC survey of the Large Magellanic Cloud: \\ Outline of the survey and initial results}

\author{Yoshifusa \textsc{Ita}$^1$ %
}
\affil{$^1$National Astronomical Observatory of Japan, 2-21-1 Osawa, Mitaka, Tokyo, 181-8588, Japan}
\email{yoshifusa.ita@nao.ac.jp}

\author{Takashi \textsc{Onaka}$^2$, Daisuke \textsc{Kato}$^2$, Toshihiko \textsc{Tanab\'{e}}$^3$, Itsuki \textsc{Sakon}$^2$, Hidehiro \textsc{Kaneda}$^4$,}
\author{Akiko \textsc{Kawamura}$^5$, Takashi \textsc{Shimonishi}$^2$, Takehiko \textsc{Wada}$^4$, Fumihiko \textsc{Usui}$^4$, Bon-Chul \textsc{Koo}$^6$,}
\author{Mikako \textsc{Matsuura}$^1$, Hidenori \textsc{Takahashi}$^7$, Yoshikazu \textsc{Nakada}$^{3}$, Tetsuo \textsc{Hasegawa}$^1$, Motohide \textsc{Tamura}$^1$,}

\affil{$^2$Department of Astronomy, Graduate School of Science, The University of Tokyo, \\ Bunkyo-ku, Tokyo 113-0033, Japan}
\affil{$^3$Institute of Astronomy, Graduate School of Science, The University of Tokyo, \\ 2-21-1 Osawa, Mitaka, Tokyo 181-0015, Japan}
\affil{$^4$Institute of Space and Astronautical Science, Japan Aerospace Exploration Agency, \\ 3-1-1 Yoshinodai, Sagamihara, Kanagawa 229-8510, Japan}
\affil{$^5$Department of Astrophysics, Nagoya University, Chikusa-ku, Nagoya 464-8602, Japan}
\affil{$^6$Department of Physics and Astronomy, Seoul National University, Seoul 151-742, Korea}
\affil{$^7$Gunma Astronomical Observatory, 6860-86 Nakayama, Takayama, Agatsuma, Gunma 377-0702, Japan}



%

\KeyWords{Galaxies:Magellanic clouds infrared:stars stars:AGB and post-AGB, YSO ISM:supernova remnants} 

\maketitle

\begin{abstract}
We observed an area of 10 deg$^2$ of the Large Magellanic Cloud using the Infrared Camera on board AKARI. The observations were carried out using five imaging filters (3, 7, 11, 15, and 24 micron) and a dispersion prism (2 -- 5 micron, $\lambda / \Delta\lambda$ $\sim$ 20) equipped in the IRC. This paper describes the outline of our survey project and presents some initial results using the imaging data that detected over 5.9$\times$10$^5$ near-infrared and 6.4$\times$10$^4$ mid-infrared point sources. The 10 $\sigma$ detection limits of our survey are about 16.5, 14.0, 12.3, 10.8, and 9.2 in Vega-magnitude at 3, 7, 11, 15, and 24 micron, respectively. The 11 and 15 micron data, which are unique to AKARI IRC, allow us to construct color-magnitude diagrams that are useful to identify stars with circumstellar dust. We found a new sequence in the color-magnitude diagram, which is attributed to red giants with luminosity fainter than that of the tip of the first red giant branch. We suggest that this sequence is likely to be related to the broad emission feature of aluminium oxide at 11.5 micron. The 11 and 15 micron data also indicate that the ([11] $-$ [15]) micron color of both oxygen-rich and carbon-rich red giants once becomes blue and then turns red again in the course of their evolution, probably due to the change in the flux ratio of the silicate or silicon carbide emission feature at 10 or 11.3 micron to the 15 micron flux.
\end{abstract}

\section{Introduction}
Owing to its proximity ($\sim$ 50 kpc; e.g., \cite{feast1987}) and face-on geometry, the Large Magellanic Cloud (LMC) is an ideal natural laboratory for the study of various astrophysical fields. It is located at a high Galactic latitude of $\sim$ $-$36$^\circ$, and we expect less contamination of foreground stars and less interstellar extinction. It is far enough to neglect its depth so that we can reasonably assume that objects in the LMC are all at the same distance from us. 
The apparent size of the LMC is also in a size, for which the entire galaxy can be surveyed in a reasonable amount of time to study the material circulation processes and star-formation history in a galactic scale.
In the meanwhile, it is close enough to resolve and study individual objects even with relatively small telescopes. Moreover, the mean metallicity of the LMC is known to be small ($\sim$ 1/4) compared to the solar, intriguing us to study the influence of low metallicity on various astrophysical phenomena.


\begin{figure}[htbp]
  \begin{center}
    \FigureFile(86mm,86mm){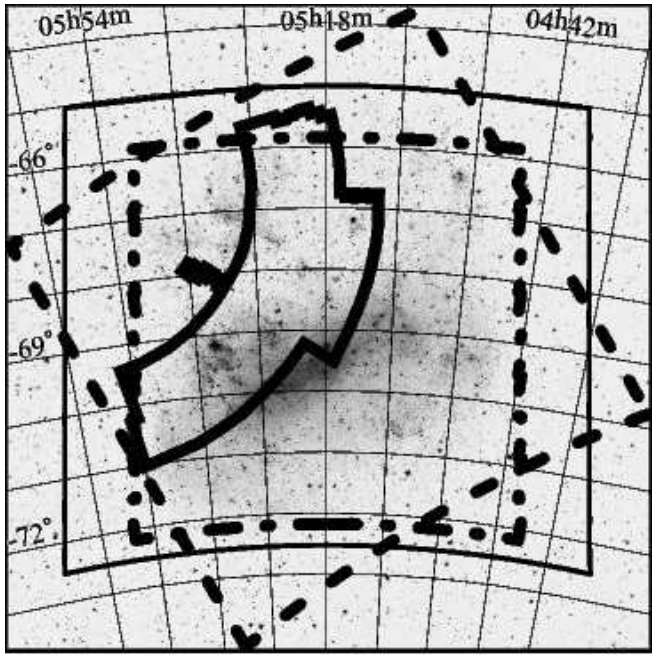}  
  \end{center}
  \caption{Observed area of the AKARI IRC survey (thick solid outline), overlaid on the photographic image kindly provided by Mr. Motonori Kamiya. The thick dashed outline indicates the coverage of the \textit{Spitzer}/SAGE survey (7$^\circ$$\times$7$^\circ$, \cite{meixner2006}), the thick dash-doted outline shows the coverage of the IRSF/SIRIUS near-infrared survey (6.3$^\circ$$\times$6.3$^\circ$, \cite{kato2007}), and the thin solid outline represents the coverage of the Magellanic clouds optical photometric survey (8.5$^\circ$$\times$7.5$^\circ$, \cite{zaritsky2004}).}
  \label{fig:surveyregion}
\end{figure}

Because of these characteristics there have been a number of survey projects of the LMC in various wavelengths (Hard X-ray: \cite{gotz2006}; Soft X-ray: e.g., \cite{long1981} with Einstein, \cite{sasaki2000}, 
\cite{haberl1999} with ROSAT, and \cite{haberl2003} with XMM; UV: \cite{smith1987}; H$\alpha$: \cite{kennicut1986}, \cite{gaustad2001}; Optical: \cite{zaritsky2004}; NIR: e.g., \cite{cioni2000a} with DENIS, \cite{nikolaev2000} with 2MASS, and \cite{kato2007} with IRSF/SIRIUS; MIR\&FIR: e.g., \cite{israel1986} with IRAS, \cite{egan2003} with MSX, and \cite{meixner2006} with \textit{Spitzer}/SAGE; [CII]: \cite{mochizuki}; CO: e.g., \cite{mizuno2001} with NANTEN; HI: \cite{luks1992}, \cite{kim1998}). 

A ground-based near-infrared (NIR) survey detected about fifteen million sources in the $\sim$ 40 deg$^2$ area of the LMC (\cite{kato2007}). On the other hand, previous mid-infrared (MIR) surveys detected only a few thousands of sources in $\sim$ 100 deg$^2$ area of the LMC (\cite{egan2003}), which is too shallow to compare with the survey observations in other wavelengths. Moreover, the angular resolutions of previous MIR surveys were not high enough to make secure cross-identifications with other survey catalogs. This situation is significantly improved with the advent of the \textit{Spitzer Space Telescope} (SST; \cite{werner2004}) and AKARI, which have instruments capable of deep mid-infrared observations with the high spatial resolution. The \textit{Spitzer} SAGE project (\cite{meixner2006}) carried out an uniform and unbiased imaging survey of about 49 deg$^2$ area of the LMC, providing photometry for about 2$\times$10$^5$ sources with all IRAC \citep{fazio2004a} bands at two epochs separated by 3 months.

We carried out near- to mid-infrared imaging and near-infrared spectroscopic survey toward the LMC with AKARI. The entire LMC was also observed as part of the All-Sky survey at 6 bands in the mid- to far-infrared with AKARI (\cite{ishihara2006}; \cite{kawada2007}). AKARI observations provide multi-band (11 bands) data of the LMC from the near- to far-infrared (FIR), which will give us a new insight into the various phenomena occurring in the LMC. In this paper, we provide an overview of our LMC survey project and present some initial results using the preliminary photometric catalog of bright point sources. 

Then we will make general analyses using the catalog with an emphasis on the unique characteristics of the IRC observations compared to {\it Spitzer} observations. The first AKARI/IRC LMC point source catalog is planned to be released to the public in 2009.

\section{The AKARI IRC LMC survey project}
The Japan Aerospace Exploration Agency launched an Infrared Satellite, ASTRO-F (\cite{murakami2007}) at 21:28 UTC on February 21st, 2006 from the Uchinoura Space Center. Once in orbit ASTRO-F was renamed "AKARI". AKARI has a 68.5 cm telescope and two scientific instruments, namely the InfraRed Camera (IRC; \cite{onaka2007}) and the Far-Infrared Surveyor (FIS; \cite{kawada2007}). Both instruments have low- to moderate-resolution spectroscopic capability. The IRC has nine imaging bands and six dispersion elements covering from 2.5 to 26 $\mu$m wavelength range (Unfortunately, one of the dispersion elements that was expected to cover 11 to 19 $\mu$m range became opaque during the ground test operation, and defunct in orbit). The FIS will observe in 4 far-infrared bands between 50 and 180 $\mu$m. One of the primary goals of the AKARI mission is to carry out an All-Sky survey with the FIS and the IRC at six bands from 9 to 180 $\mu$m (\cite{ishihara2006}; \cite{kawada2007}). As a result, the entire LMC has been mapped in 9, 18, 65, 90, 140, and 160 $\mu$m wavebands. The 5 $\sigma$ detection limits for one scan observation of the All-Sky survey are shown in Figure~\ref{fig:sensitivity}. As described below, the LMC is located at a relatively high visibility region for AKARI's orbit. AKARI scanned LMC several times and thus better detection limits should be obtained by further data reduction of the All-Sky survey data. The All-Sky survey observations of the LMC will be described elsewhere and will not be discussed in this paper.

\begin{figure}[htbp]
  \begin{center}
    \FigureFile(85mm,85mm){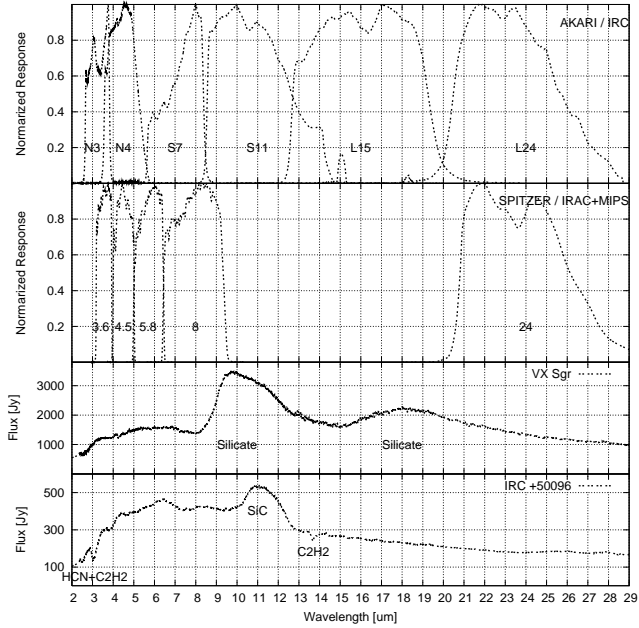}
  \end{center}
  \caption{The normalized spectral response function of AKARI IRC bands and \textit{Spitzer} IRAC and MIPS bands. As references, the ISO SWS spectra of two representative Galactic AGB stars (VX Sgr as O-rich AGB and IRC+50096 as C-rich AGB) with circumstellar dust features are shown.}
  \label{fig:filters}
\end{figure}

In addition to the All-Sky survey in mid- and far-infrared, AKARI carried out two large-area legacy surveys (LS) in pointing mode. The LMC survey project (PI. T.Onaka) is one of the two LS programs. The other is the North Ecliptic Pole survey project (PI. H.Matsuhara; \cite{matsuhara2007}). These survey areas are located at high ecliptic latitudes, where the visibility is high for AKARI's sun-synchronous polar orbit. Therefore a batch of observing time can be allocated for pointing observations in these areas after allocating scan paths for the All-Sky survey. We used the IRC to make imaging and spectroscopic mapping observations of the main part of the LMC.

The IRC is comprised of three independent channels; NIR, MIRS, and MIRL, which cover the 1.8 $-$ 5.5 $\mu$m, 4.6 $-$ 13.4 $\mu$m, and 12.6 $-$ 26.5 $\mu$m wavelength range, respectively. Each of the three channels has three imaging bands and two dispersion elements, which can be switched during a pointed observation opportunity. Each of the channels has a wide field of view (FOV) of about \timeform{10'} $\times$ \timeform{10'}, suitable to survey observations. Compared to the contemporary SST the AKARI IRC has the following characteristics.
\begin{itemize}
\item The IRC's imaging filters cover the wavelength range continuously from 2.5 to 26 $\mu$m. In particular, it has 11$\mu$m (S11) and 15$\mu$m (L15) imaging bands, which fill the gap between the IRAC \citep{fazio2004a} and MIPS \citep{rieke2004} on board the SST. The wavelength range that S11 and L15 bands cover contains interesting spectral features such as the silicate 10 and 18 $\mu$m bands. Figure~\ref{fig:filters} shows the spectral response functions of the IRC, IRAC, and MIPS bands, together with ISO-SWS spectra (\cite{sloan2003}) of an oxygen-rich AGB star (VX Sgr) and a carbon-rich AGB star (IRC $+$50096). The IRC S11 and L15 probe the silicate bands efficiently.
\item One of the most significant features of the IRC is its ability to perform slit-less spectroscopy in addition to imaging. It can simultaneously obtain near- to mid-infrared (about 2 $-$ 13 $\mu$m) continuous spectra of all sources present in its large FOV of about \timeform{10'} $\times$ \timeform{10'} within a pointing opportunity.
\end{itemize}
These features make AKARI IRC unique and complementary to IRAC and MIPS on the SST. Refer to \citet{onaka2007} for instrumental details and imaging performance of the IRC. General information on the IRC spectroscopic mode, particularly on the slit-less spectroscopy, is given in \citet{ohyama2007}.

\section{Scientific objectives}
Interstellar matter in a galaxy is thought to evolve through various processes. Stars are born in interstellar clouds by accreting the interstellar matter and then appear as young stellar objects (YSOs), such as T Tauri stars or Herbig Ae/Be stars. Low- and intermediate-mass stars evolve into white dwarfs after spending most of their lives in the main sequence. During the course of their evolution to white dwarfs, they lose mass through so called mass-loss process and supply dust grains and gas to the interstellar space. High-mass stars end their lives as supernovae and enrich the interstellar medium with a large amount of nuclear synthesized elements.
They can also trigger the formation of next generation stars through their strong UV photons and stellar winds (\cite{hosokawa2006}; \cite{koo2008}).
Interstellar grains play an important role in every aspect of these interstellar processes, including the star-formation. All of these phenomena can be studied most effectively in the infrared. The \textit{Spitzer} SAGE project \citep{meixner2006} have provided IRAC (3.5, 4.5, 5.8, and 8.0 $\mu$m) and MIPS (24, 70, and 160 $\mu$m) images for a wide area (7$^\circ$ $\times$ 7$^\circ$) of the LMC. Adding to the SAGE observation, the complementary AKARI IRC (3.2, 7.0, 9.0, 11.0, 15.0, and 18.0 $\mu$m) and FIS (65, 90, 140, and 160 $\mu$m) observations of the LMC will provide an opportunity to make a thorough study of these material circulation processes and local star-formation history of the LMC in a galactic scale owing to its wide spectral coverage and high sensitivities together with the sufficient spatial resolution. 

\begin{figure}[htbp]
  \begin{center}
   \FigureFile(88,88mm){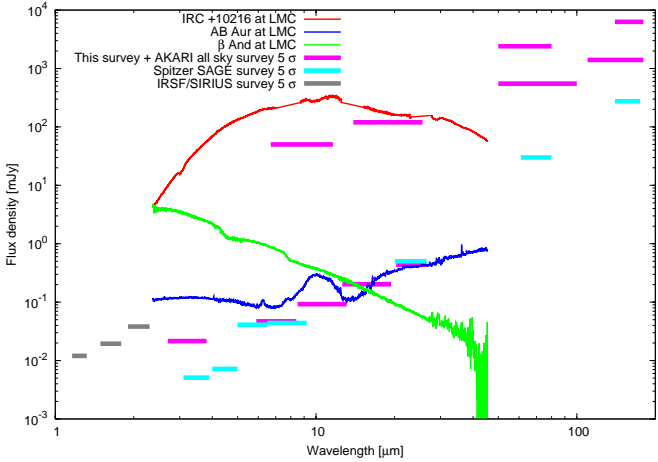}
  \end{center}
  \caption{A graphic representation of the 5 $\sigma$ detection limits of the AKARI LMC survey and the All-Sky survey (magenta). For comparison, the 5 $\sigma$ detection limits of the \textit{Spitzer} SAGE survey and the IRSF/SIRIUS survey are also shown (cyan and gray, respectively). Some ISO SWS spectra (\cite{sloan2003}) of famous galactic stars are scaled to fit the distance to the LMC and included in the figure, as examples of Herbig Ae/Be stars (AB Aur), red giants with luminosities below the tip of the first red giant branch ($\beta$ And), and dusty red giants (IRC +10216) in the LMC.}
  \label{fig:sensitivity}
\end{figure}

Figure~\ref{fig:sensitivity} is a diagrammatic representation of 5 $\sigma$ detection limits of our survey, \textit{Spitzer} SAGE survey, and the IRSF/SIRIUS near-infrared survey. The 5 $\sigma$ detection limits of our survey are calculated by scaling the 10 $\sigma$ detection limits in Table~\ref{table:survey} (see section 5.4.1). We take the 5 $\sigma$ detection limits for the SAGE survey from the "SAGE Data Description: Delivery 1"\footnote{http://sage.stsci.edu/index.php}. The detection limits are expected to be further improved in the
final catalog deliveries (\cite{meixner2006}). It is clear that the detection limits of our survey are comparable to that of the \textit{Spitzer} SAGE survey. The ISO SWS spectra of galactic famous stars of known distances (the distances were taken from \cite{crosas1997} for IRC +10216, and \cite{perryman1997} for others) are included for illustrative purposes after scaling their fluxes at the distance of the LMC. We assumed a distance modulus of 18.5 mag to the LMC. The figure indicates that all of red giants above the tip of the first red giant branch, and a large fraction of Herbig Ae/Be stars in the LMC are detected in our survey.

\subsection{Star-formation and young stars}
The LMC contains a class of so called populous clusters (e.g., \cite{kumai1993}) that have no Galactic counterpart. They have masses intermediate between Galactic globular clusters and open clusters. Characteristic environments in the LMC, such as low mean metallicity (about 1/4 of solar vicinity), low dust to gas ratio (about 1/4 of Milky Way; \cite{koornneef1982}) and strong ultraviolet interstellar radiation (e.g., \cite{cox2006}), may play a key role in the star-formation process of populous clusters (\cite{fukui2005}). Detailed studies of the star-formation in the LMC will provide insights into these processes.
The IRAS survey has detected protostars in the LMC only brighter than 10$^4$ $\sim$ 10$^5$ L$_\odot$ (e.g., \cite{vanloon2005}). Although only the brightest end of the YSOs in the LMC can be detected by the IRC (see the detection limits provided in Table~\ref{table:survey}), the IRC imaging data provide us with a valuable dataset to investigate the nature of such objects from the spectral energy distribution from near- to mid-infrared as well as the water ice feature at 3 $\mu$m and/or silicate absorption at 10 and 20 $\mu$m. AKARI observations of the LMC enable us to probe the luminosity function of protostars down to about 10 L$\odot$.

The detection limits of IRC bands are not deep enough to detect T Tauri stars in the LMC. However the IRC observations can nearly completely detect young stars of intermediate mass (Herbig Ae/Be) as well as classical Be stars. It is difficult to separate Herbig Ae/Be stars from classical Be stars and another kind of young stars with different masses and ages, based on near-infrared data alone. Mid-infrared data provided by the IRC make clear identification of these objects. AKARI IRC observations are able to provide the initial mass function of objects later than the Class 1 phase for M $>$ 2 -- 3 M$_\odot$ except for dense clusters, where the source confusion limits the detection. They allow us to compare the spatial distribution of YSOs with that of CO clouds in a statistical way and investigate the star-formation history in the LMC. Particularly interesting is the search for embedded star clusters in "clusterless" giant molecular clouds (GMCs), such as N159S (\cite{bolatto2000}) and investigation of whether or not these clusterless GMCs are really not associated with star clusters. The 5 band IRC observations together with 4 band FIS All-Sky survey data allow us to distinguish these YSOs unambiguously and study the nature of detected objects in detail. 

\subsection{Evolved stars}
Stars of low- to intermediate-mass (stars with main sequence masses M$_{\textrm{ms}}$ of about 0.8 -- 8 M$_\odot$) eventually evolve into white dwarfs. The mass M$_{\textrm{wd}}$ distribution of white dwarfs peaks around 0.6 M$_\odot$ (e.g., \cite{kepler2007}). Therefore, the difference, $\Delta M = (M_{\textrm{ms}} - M_{\textrm{wd}})$ should be lost during their life in the mass-loss process, which enriches gas and dust in the interstellar medium. It is observationally known that the maximum mass-loss rate is attained at the latest stage of their Asymptotic Giant Branch (AGB) evolutionary phase, where stellar pulsation is usually associated. However, the onset and evolution of the mass-loss process and its relation to pulsation are not yet completely understood. 

As described above, the LMC has a number of young clusters, which also provide a unique place to study the late evolutional stage of intermediate-mass stars. Star clusters of intermediate ages still contain a number of AGB stars that are losing mass. Since their ages and masses are well determined, the study of mass-losing AGB stars in the clusters can elucidate the mass-loss process for stars with given ages and masses quantitatively for the first time. 

\citet{blum2006} used Spitzer SAGE survey epoch 1 catalog to discuss and identify the evolved stars in the infrared color-magnitude diagrams. Ita et al. (2004a,b) studied OGLE variables in the Magellanic Clouds (\cite{udalski1997}; \cite{zebrun2001}). They detected most of optical variable stars in the central parts of the Large and Small Magellanic Clouds. The association of the IRC data with these variability catalogs enables us to investigate the mass-loss phenomena in terms of the pulsation activity directly. 
The IRC observations can detect all mass-losing AGB stars in the LMC. It should be emphasized that the IRC band S11, L15, and L24 are particularly important to determine the chemistry of the envelope (O-rich or C-rich) and the mass-loss rate accurately because of the presence of the silicate bands at 10 and 18 $\mu$m and the silicon carbide band at 11.3 $\mu$m.

\subsection{Supernova remnants}
Supernovae (SNe) are believed to be a major source of interstellar dust since the expected
supply rate from late-type stars is too short to balance with the dust destruction rate in the
interstellar space (Jones et al. 1994; 1996). The role of the dust production in SNe is crucial to understand the circulation and evolution of materials in the interstellar medium (\cite{dwek1998}). However, to date, no clear observational evidence has been given for the large production of dust grains from SNe (\cite{meikle2007}). The condensation efficiency could be either very low (\cite{clayton2001}) or moderate (\cite{todini2001}). A study of Cas A SNR by \textit{Spitzer} clearly demonstrates the observational difficulty of dust grains associated with SNRs in our Galaxy (\cite{krause2004}). SNRs in the Galaxy are easily confused by the material on the line of sight because most of them are located on the Galactic plane. In this regard, the LMC provides an ideal opportunity to study material ejected in SNRs without suffering from serious confusion problems. Recent studies of SNRs in the LMC with Spitzer and AKARI provide
significant information on the origin of the IR emission in SNRs and the interstellar dust lifecycle in terms of dust formation by SNe and dust distruction by SNR blast waves (e.g., \cite{borkowski2006}; \cite{bwilliams2006}; \cite{rwilliams2006}; \cite{rho2008}; \cite{seok2008}).


The interaction of SNRs with their surroundings is also an important phenomenon, providing opportunities for studying strong shocks and sometimes triggered star-formation (Koo et al. 2007; 2008).
However, the observational study of the interaction is also hampered by confusion for Galactic SNRs. There are more than 40 SNRs known in the LMC (\cite{filipovic1998}; \cite{haberl1999}), and only 10 of them have been detected by IRAS. SNRs appear very patchy in the infrared (\cite{douvion2001}) and their spectral energy distribution indicates the multi-temperature nature. They also have many emission lines, which could affect the broad-band photometry. Strong emission lines expected in the MIR are [ArII] 7.0 $\mu$m, [NeII] 12.8 $\mu$m, [NeIII] 15.6 $\mu$m, [FeII] 17.9 $\mu$m, [FeII] 24.5, 26.0 $\mu$m, and [SiII] 34.8 $\mu$m (e.g., \cite{arendt1999}; \cite{temim2006}; \cite{rho2008}). Dust grains in SNRs may be transiently heated by high-energy electron hits, which results in a wide dust temperature distribution (\cite{dwek1986}). A peculiar band may also be present in the MIR (\cite{arendt1999}), which might be associated with dust grains formed in the SNR (\cite{chan2000}; \cite{onaka2008}). Therefore, multi-band photometry spanning over a wide spectral range with sufficient spatial resolution is crucial to study the nature of the infrared emission from SNRs and could confirm the physical association of the emission with SNRs.

\subsection{Interstellar Matter}
Diffuse infrared emission from interstellar dust consists of several components: thermal emission in the FIR from submicron size grains, excess continuum emission in 15.80 $\mu$m, and the unidentified infrared (UIR) emission bands arising from aromatic materials (\cite{onaka2000}). Each component should play a significant role in the interstellar process and their relative variations in abundance should provide valuable information on the physical conditions of the interstellar medium. However, little has been known for formation, evolution (processing), and destruction of each dust component. A study of the UIR bands in our Galaxy has suggested possible variations in the relative abundance to the submicron grains as well as in the band appearance with the Galacto-centric distance for the first time (\cite{sakon2004}). Observations of the Galactic plane pick up diffuse emissions from every component on the line of sight and it is impossible to separate out the emission from a particular region. The LMC again provides an ideal place to study the diffuse infrared emission from dust grains in several different environments compared to our Galaxy because of its nearly face-on geometry. It has been suggested that the nature of the infrared emission from the LMC is different from our Galaxy (\cite{aguirre2003}), which may be related to the low metallicity of the LMC. 
A recent study of IRAS data has indicated that the color of the IRAS 12 $\mu$m to 25 $\mu$m intensities seems to be high in CO clouds without young star clusters, whereas it is low in regions with YSOs (\cite{sakon2005}). There is also an indication that the 12 $\mu$m to 100 $\mu$m ratio is different inside the supergiant shells (\cite{kim1998}), which is also related to the star-formation activity (\cite{yamaguchi2001}). 
Since the IRAS 12 $\mu$m band intensity includes both the UIR and the MIR excess emissions, it is important to separate the contribution from each component and compare it directly to the FIR component. 
The IRC S7 filter is most sensitive to the 6.2 and 7.7 $\mu$m UIR bands, whereas the S11 band will probe the UIR 11.2 $\mu$m band. The S7 to S11 band ratio together with the MIR and FIR All-Sky survey data can investigate variations in the relative band strength efficiently. The L15 and L24 band data determine the contribution and spectral profile of the MIR excess component unaffected by the UIR bands.

\begin{figure}[htbp]
  \begin{center}
   \FigureFile(85,85mm){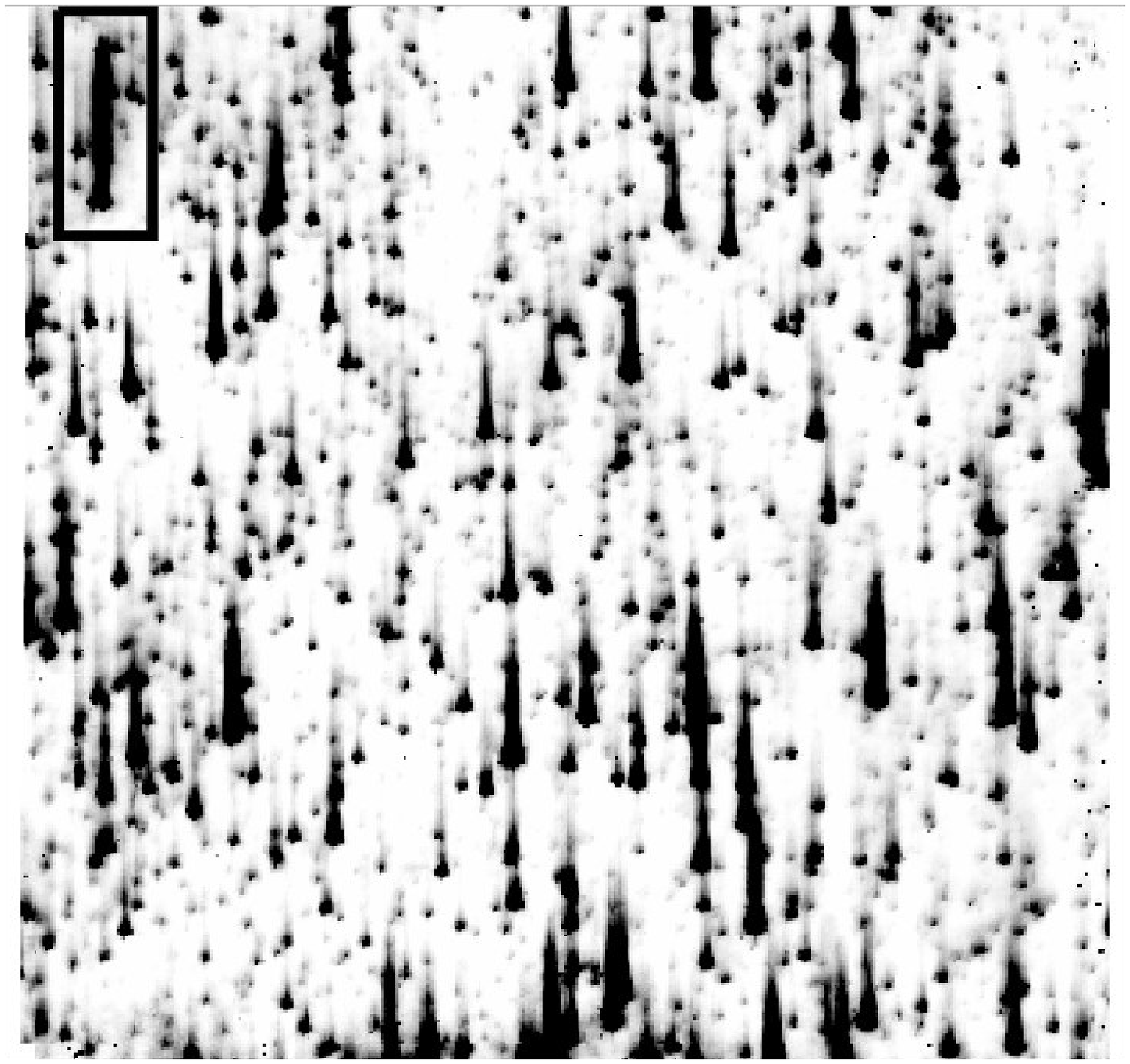}
   \FigureFile(85,85mm){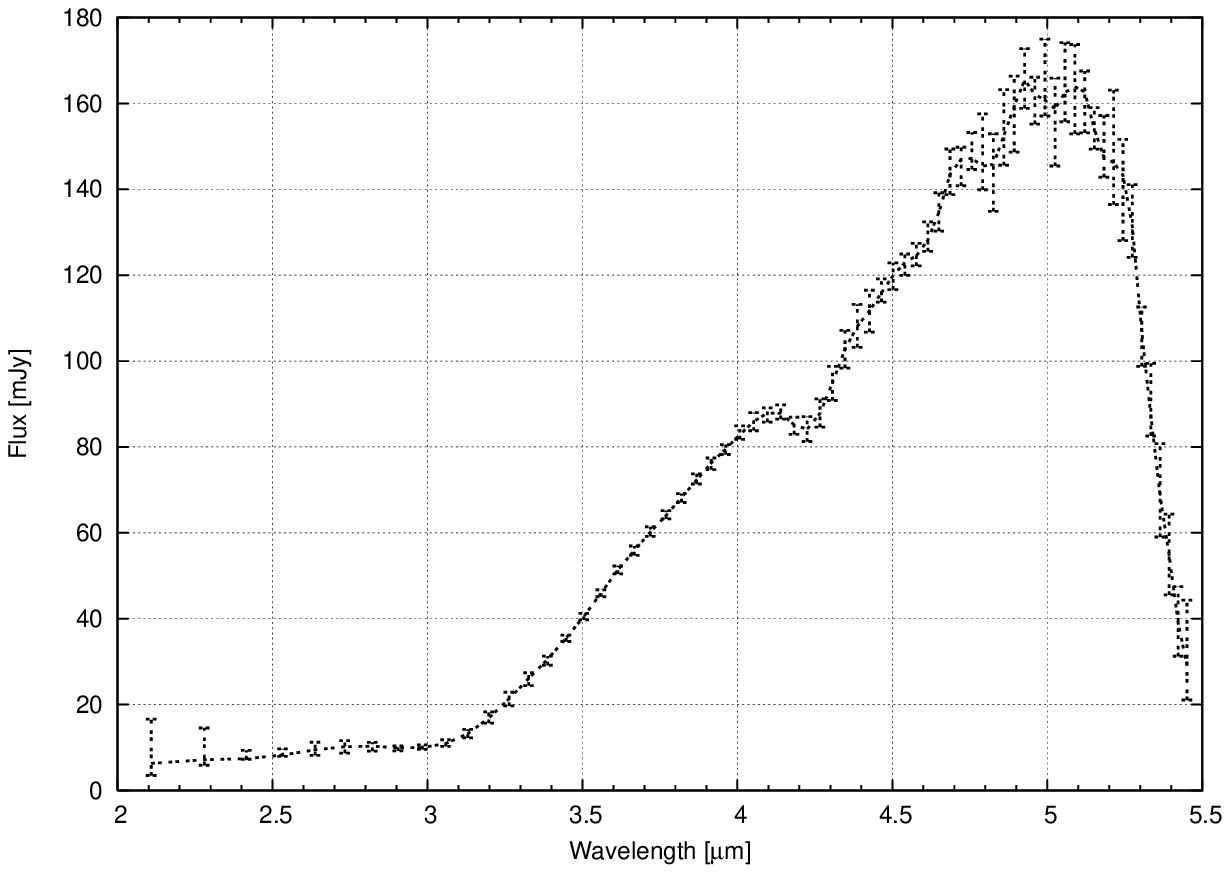}
  \end{center}
  \caption{A sample image of NP slit-less spectroscopy data taken in our survey and a reduced NP spectrum of a source indicated in the box.}
  \label{fig:samplespecimage}
\end{figure}

\section{Observations}
\subsection{IRC astronomical observing template and pointing observation sequence}
To cover a wide range of the spectral energy distribution of a celestial body, we use the AKARI IRC AOT02 observing template with a special option prepared only for the LMC survey. The AOT02 with the special option yields not only imaging data at 3 (N3), 7 (S7), 11 (S11), 15 (L15), and 24 $\mu$m (L24), but also low resolution ($\lambda / \Delta \lambda$ $\sim$ 20) 2.5 $-$ 5 $\mu$m (NP) spectral data at three dithered sky positions in a pointing opportunity. Figure~\ref{fig:samplespecimage} shows an example of the NP spectroscopic data taken in our survey. The reduced spectrum of a red source indicated in the box is also shown in the figure. 
Discussions on spectroscopic data will be given in a future paper.

As described in the IRC data user's manual (\cite{lorente2007}) the IRC observation produces short- and long-exposure data together.
The long to short exposure time ratios in each channel are 9.5, 28, and 28 for NIR, MIRS, and MIRL, respectively. It enhances the dynamic range by combining photometric results of the two different exposure-time data. In this paper, we use the combined data of long- and short-exposure for N3, but only use long-exposure data for MIRS and MIRL.

\subsection{Mapping strategy}
Due to the focal-plane layout of AKARI, the NIR and MIRS channels share the same field of view (i.e., they observe the same sky position), but the MIRL channel observes a different sky position that is offset by 25 arcmin from the NIR/MIRS center in the direction perpendicular to the AKARI's orbit. AKARI is in a sun-synchronous polar orbit at 700 km attitude along the twilight zone, and its orbit is approximately parallel to the ecliptic meridian lines with approximately 4.1 arcmin spacing at the ecliptic plane between successive orbits. We used this feature to map the LMC effectively. Each of the IRC channels has a field of view of \timeform{10'} $\times$ \timeform{10'} and adjacent fields are designed to overlap by $\sim$ \timeform{1.5'}. Since AKARI follows the Earth's yearly round, the telescope points at the same celestial position every six months, but the direction of the satellite movement relative to the position is rotated by 180 degrees. Thus the position of the MIRL channel relative to the NIR and MIRS channels is also rotated by 180 degrees in projection on the sky. Therefore, we divide the observing area into several parts to maximize the observing efficiency to map a large part of the LMC with all three channels of the IRC. Observations were carried out in three separate seasons, from 6th May 2006 to 8th June 2006, from 2nd October 2006 to 31st December 2006, and from 24th March 2007 to 2nd July 2007. Over 600 pointing observations were devoted for this project, yielding about a 10 deg$^2$ imaging and spectroscopic map of the main part of the LMC.

\section{Data reduction}
\subsection{Pipeline processing}
Raw imaging data were processed with the IRC imaging toolkit, version 20071017 (see IRC Data User's Manual \cite{lorente2007} for details). In a pointing observation using IRC, one short-exposure and three long-exposure dark data are taken at the beginning (pre-dark) and the end of the operation for all of the three IRC channels. We made long-exposure dark images for MIRS and MIRL channels by averaging over three long exposure dark data. We refer to these averaged pre-dark images as the selfdark. We subtracted selfdarks from MIRS and MIRL long-exposure images and the pipeline default super-dark from the others.

As described above, we use the AKARI IRC AOT02 observing template that produces data at three different sky positions in a pointed observation. The data in each band are spatially aligned and then coadded by taking their median to eliminate array anomalies such as bad, dead, or hot pixels, cosmic rays, and so on. The resultant coadded images have pixel sizes of \timeform{1.446''} pixel$^{-1}$, \timeform{2.340''} pixel$^{-1}$ and \timeform{2.384''} pixel$^{-1}$ for NIR, MIRS, and MIRL images, respectively. 

\subsection{Astrometry}
After the coaddition process, we calculate the coordinate transform matrix that relates the image pixel coordinates to the sky coordinates by matching detected point sources with the Two Micron All Sky Survey (2MASS; \cite{skrutskie2006}) catalog. We use at least five matched point sources for the calculation. If matching with the 2MASS catalog is unsuccessful (such cases usually occur in L15 and L24 images), then we use the SAGE point source catalog (\cite{meixner2006}) as the positional reference. The root-mean-squares of the residuals between the input 2MASS/SAGE catalog coordinates and the fitted coordinates are smaller than \timeform{1.2''}, \timeform{2.6''}, and \timeform{2.9''}, for NIR, MIRS, and MIRL images, respectively. The coordinates of AKARI sources should be accurate to that extent relative to the 2MASS and SAGE catalog coordinates.

\subsection{Photometry}
Photometry is carried out for each coadded image independently. To derive calibrated fluxes for each detected source, point spread function (PSF) fitting photometry was performed on the coadded images with the IRAF\footnote{IRAF is distributed by the National Optical Astronomy Observatories, which are operated by the Association of Universities for Research in Astronomy, Inc., under cooperative agreement with the National Science Foundation.} package DAOPHOT. We develop PSF fitting photometry software, which is similar to the one that the \textit{Spitzer} GLIMPSE\footnote{Visit http://www.astro.wisc.edu/glimpse/, and see a document "Description of Point Source Photometry Steps Used by GLIMPSE", written by Dr. Brian L. Babler.} (\cite{benjamin2003}) team uses. Our photometric process involves the following steps:
\begin{enumerate}
 \item Source extractor \citep{bertin1996} is used for each coadded image to extract sources whose fluxes are at least 3 $\sigma$ above the background. The saturated sources should not be extracted, since the central pixels of them are already masked by the IRC imaging toolkit. Even if they are detected by the source extractor, the PSF fitting does not work well for them, and they are not included in the following discussions.
 \item Aperture photometry is performed on all of the sources found in step 1, using the task PHOT with aperture radii of 10.0 and 7.5 pixels for NIR and MIRS/MIRL images, respectively. We use the same aperture radii as have been used in the standard star flux calibration (Tanab\'{e} et al. in preparation), and the aperture corrections are not applied. The resultant arbitrary fluxes are converted into the physical units by using the IRC flux calibration constants version 20071112. Then the calibrated fluxes are converted into the magnitudes of the IRC-Vega system by using the zero magnitude fluxes listed in Table~\ref{table:survey}. The offsets between the arbitrary and the calibrated magnitudes are constants. 
 \item Several point sources with moderate flux (i.e., with a good signal-to-noise ratio and unsaturated) and without other sources within 7 pixels are selected from the results of step 2. At least 8 such "good" stars are selected in the NIR/MIRS (i.e., N3, N4, S7, and S11) images, and more than 5 good stars in the MIRL (i.e., L15 and L24) images. They are used for the aperture correction to the PSF fitting photometry in the following process. 
 \item Since the shapes of the PSFs measurably vary from pointing to pointing in NIR images possibly due to jitters in the satellite pointing, the selected good stars in step 3 are used to construct a model PSF for each image. 
 We use DAOPHOT to choose the best fitting function by trying several different types of fitting functions. In MIRS/MIRL images, we use a "grand-PSF" for each band, which was built in advance. These grand-PSFs were made from "good" stars free from diffuse emission, carefully chosen by eyes from hundreds of images taken in the AKARI/IRC LMC survey data and other related ones.
 \item The PSF fitting photometry is performed on all of the sources found in the coadded images using ALLSTAR. 
The PSF does not vary appreciably over the field of view. Hence a constant PSF is assumed over an image. This PSF fitting operation is iterated, in just the same way as the GLIMPSE team does. 
 \item Aperture correction is applied to the fit magnitudes by comparing them with the aperture magnitudes of good stars, which are selected in step 3. Then, the offset value obtained in step 2 is added to the aperture corrected fit magnitudes to derive calibrated fit magnitudes.
\end{enumerate}

As indicated above, we use the zero magnitude fluxes (Tanab\'{e} et al. in preparation) to convert the photometric fluxes into the Vega magnitudes. These zero magnitude fluxes are tabulated in Table~\ref{table:survey}. It should be noted that the IRC absolute flux calibration assumes a spectral energy distribution (SED) of $f_\lambda \propto \lambda^{-1}$ or $f_\nu \propto \nu^{-1}$. Note that throughout this paper, we do not apply color corrections to the calibrated fluxes and magnitudes. No dereddening are applied either.

\begin{table*}[htbp]
  \caption{Survey properties}\label{table:survey}
  \begin{center}
    \begin{tabular}{lrrrrrr}
    \hline
    \multicolumn{1}{c}{Properties} & \multicolumn{6}{c}{IRC bands} \\ 
    \multicolumn{1}{c}{} & \multicolumn{1}{r}{NP} & \multicolumn{1}{r}{N3}  & \multicolumn{1}{r}{S7} & \multicolumn{1}{r}{S11} & \multicolumn{1}{r}{L15} & \multicolumn{1}{r}{L24} \\
    \hline
    Channel               & NIR & NIR & MIRS & MIRS & MIRL & MIRL \\ 
    Bandpass [$\mu$m]             & 1.8 - 5.5 & 2.7 - 3.8 & 5.9 - 8.4 & 8.5 - 13.1 & 12.6 - 19.4 & 20.3 - 26.5 \\ 
    Reference wavelength [$\mu$m] & -- & 3.2 & 7.0 & 11.0 & 15.0 & 24.0 \\ 
    Pixel field of view [\timeform{''}/pixel] & -- & \multicolumn{1}{r}{1.446} & \multicolumn{1}{r}{2.340} & \multicolumn{1}{r}{2.340} & \multicolumn{1}{r}{2.384} & \multicolumn{1}{r}{2.384} \\
    Dispersion [$\mu$m/pixel] & 0.06 at 3.5 $\mu$m& -- & -- & -- & -- & -- \\
    Exposure time: long exposure\footnotemark[1] [s] & 133.2432 & 133.2432 & 147.2688 & 147.2688 & 147.2688 & 147.2688 \\
    Exposure time: short exposure\footnotemark[1] [s] & 14.0256 & 14.0256 & 1.7532 & 1.7532 & 1.7532 & 1.7532 \\
    10 $\sigma$ detection limit\footnotemark[2] [mJy] & -- & 0.086 & 0.188 & 0.369 & 0.811 & 1.744 \\
    10 $\sigma$ detection limit\footnotemark[2] [mag] & -- & 16.50 & 14.00 & 12.54 & 10.74 & 9.16 \\
    Saturation limit\footnotemark[2,3] [mJy] & 12500 at 3.5 $\mu$m & 250  & 1800   & 1800   & 2500   & 23000  \\
    Saturation limit\footnotemark[2,3] [mag] & 3.6 at 3.5 $\mu$m   & 7.8    & 4.0    & 3.3    & 2.0    & -1.1   \\
    Zero magnitude flux\footnotemark[4] [Jy]                 & --    & 343.34 & 74.956 & 38.258 & 16.034 & 8.0459 \\
    Number of detected sources      & -- & $>$5.9$\times$10$^5$ & $>$8.8$\times$10$^4$ & $>$6.4$\times$10$^4$ & $>$2.8$\times$10$^4$ & $>$1.5$\times$10$^4$ \\
    \hline
    \end{tabular}
    \end{center}
    \footnotemark[1] These are the most common values. Total exposure time per pixel can be different from coadded images to images, because some data badly damaged by cosmic-ray and/or hitting of charged particles are discarded. \\
    \footnotemark[2] For point sources. \\
    \footnotemark[3] Numbers are taken from ASTRO-F Observer's Manual version 3.2 (\cite{afobsman2005}). \\
    \footnotemark[4] Tanab\'{e} et al. in preparation.
\end{table*}

\subsection{Compilation of a preliminary catalog}
We carried out photometry on each coadded image as outlined in the previous section. It produced photometric catalogs for each band for individual images. At first, the catalogs for individual IRC imaging bands (i.e., N3, S7, S11, L15, and L24) are made. Because of the field overlap, there should be substantial multiple entries of a single source in the catalogs. We eliminate such multiple entries based on the spatial proximity ($|\Delta r| \le$ \timeform{1.5''}). We adopt the one with a better S/N and discarded the other(s). We use both long- and short-exposure data of N3, and we merged them using a positional tolerance of \timeform{1.5''}. If a source is detected in both short and long exposure data, we adopt the one with a better S/N and discard the other. Then, two corresponding band catalogs are merged to make individual IRC channel catalogs. The, S7 and S11 (L15 and L24) catalogs are merged using a positional tolerance of \timeform{1.5''} (\timeform{1.5''}) to make a MIRS (MIRL) channel catalog. For the matched sources, their coordinates are recalculated by taking an average of the coordinates of each band. In rare cases that more than two sources are present within the tolerance radius, the closest one is always adopted and the others were listed as solitary sources. The resultant individual channel catalogs are further merged to make a grand catalog using a positional tolerance of \timeform{1.5''}.


\subsubsection{Detection limits and completeness}
\begin{figure}[htbp]
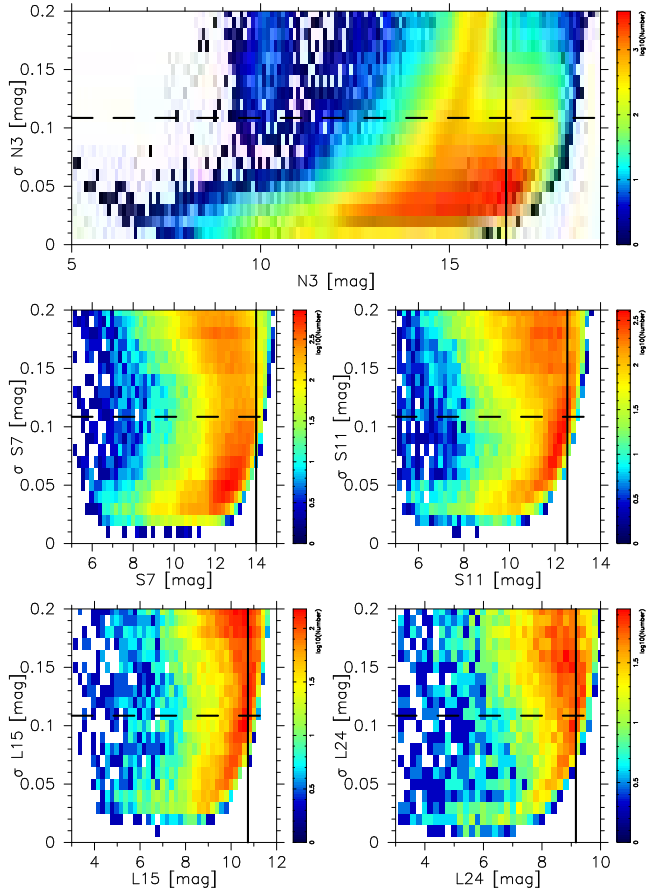

  \begin{center}
   \FigureFile(85,85mm){figure5.ps}
  \end{center}
  \caption{Photometric uncertainties as a function of the magnitude at each IRC band. The sizes of the bins are 0.1 $\times$ 0.01 mag for N3, and 0.2 $\times$ 0.01 mag for S7, S11, L15, and L24. The dashed lines show 10 $\sigma$ errors, and the solid lines indicates the 10 $\sigma$ detection limits: 16.50, 14.00, 12.54, 10.74, and 9.16 mag in N3, S7, S11, L15, and L24, respectively.}
  \label{fig:error}
\end{figure}

The distributions of the photometric uncertainty versus the magnitude in the IRC bands are shown in Figure~\ref{fig:error}. The photometric uncertainty includes errors in the fitted magnitude calculated by IRAF/ALLSTAR, in the aperture correction factor, and also in the ADU-to-Jy conversion factor. The horizontal dashed lines show the signal to noise ratio (S/N) of 10, and the vertical solid lines show the 10 $\sigma$ detection limits, which are defined as the faintest magnitudes at which the mode of photometric uncertainties of the sources in 0.02 mag bins exceed 10 $\sigma$, except for the N3 data. In the calculation of the 10 $\sigma$ detection limit for the N3 data, we use only data detected in long exposure images, and take a mean of photometric uncertainties in 0.02 mag bins instead of taking a mode of them. Visual inspections showed that sources can be detected only in short exposure images, and this is why there is a sequence around 15 mag in the N3 error plot. There seems to be several causes to explain this. For example, not a small number of long exposure images are damaged by column pulldowns caused by saturated stars, and sources in such polluted columns are not detected in the long exposure images. The estimated 10 $\sigma$ detection limits are 16.50, 14.00, 12.54, 10.74, and 9.16 in N3, S7, S11, L15, and L24, respectively. These detection limits are comparable to that of the \textit{Spitzer} SAGE survey (see Figure~\ref{fig:sensitivity}).

\begin{figure}[htbp]
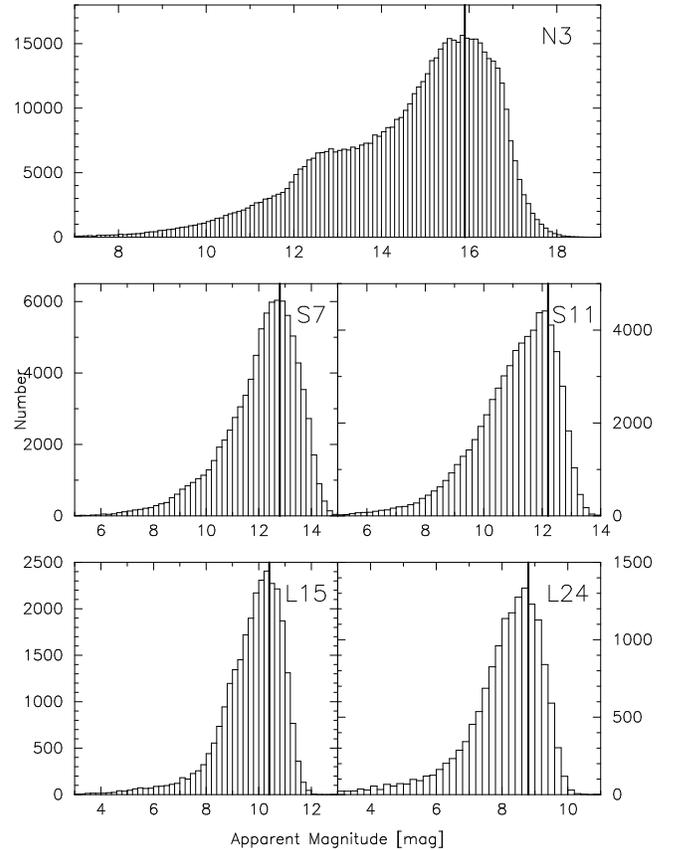

  \begin{center}
   \FigureFile(85,85mm){figure6rv.ps}
  \end{center}
  \caption{Magnitude distribution of the sources in the AKARI LMC survey catalog. The sizes of the bins are 0.1 mag for N3, and 0.2 mag for S7, S11, L15, and L24. The solid lines show the peak of the source count histograms, below which the photometry is incomplete: 15.9, 12.8, 12.2, 10.4, and 8.8 mag in N3, S7, S11, L15, and L24, respectively.}
  \label{fig:magdist}
\end{figure}

The magnitude distributions of the sources in the preliminary catalog are shown in Figure~\ref{fig:magdist}. The sizes of the bins are 0.1 mag for N3 and 0.2 mag for S7, S11, L15, and L24. The vertical solid lines show peak magnitudes, below which photometry is incomplete. The peak magnitudes are 15.9, 12.8, 12.2, 10.4, and 8.8 mag in N3, S7, S11, L15, and L24, respectively.


The 10 $\sigma$ detection limits for point sources and source counts in each IRC band are summarized in Table~\ref{table:survey} together with other properties of our survey.

\subsubsection{Cross-identifications with the existing point source catalogs}
We cross-identified our catalog with the following existing catalogs using a positional tolerance of \timeform{1.5''}, and use the result for discussion in the rest of the paper. 
\begin{itemize}
\item The Magellanic Clouds Photometric Survey catalog (\cite{zaritsky2004}): It lists U, B, V, and I stellar photometry of the central 64 deg$^2$ area of the LMC.
\item The IRSF Magellanic Clouds Point Source Catalog (\cite{kato2007}): The IRSF catalog lists JHK$_s$ photometry of over 1.4$\times$10$^7$ sources in the central 40 deg$^2$ area of the LMC. Compared to the contemporary DENIS (\cite{cioni2000a}) and 2MASS (\cite{skrutskie2006}) catalogs, the IRSF catalog is more than two mag deeper at K$_s$-band and about four times finer in spatial resolution. 
\item The \textit{Spitzer} SAGE catalog (\cite{meixner2006}): It lists near- to far-infrared photometry of sources in the central 49 deg$^2$ area of the LMC. 
\end{itemize}
The observed regions of the above existing catalogs are indicated in Figure~\ref{fig:surveyregion}. The figure shows that a part of the AKARI IRC survey region is outside the bounds of the IRSF/SIRIUS survey area. We do not use the AKARI sources located outside the IRSF/SIRIUS area in the following discussions. Throughout this paper, the numbers bracketed by $[~]$ designates the data of the SAGE catalog, for example, [3.6] indicates the photometry in the IRAC 3.6 $\mu$m band.

\subsubsection{Comparison to the \textit{Spitzer} SAGE survey catalog}
\begin{figure}[htbp]
  \begin{center}
   \FigureFile(85,85mm){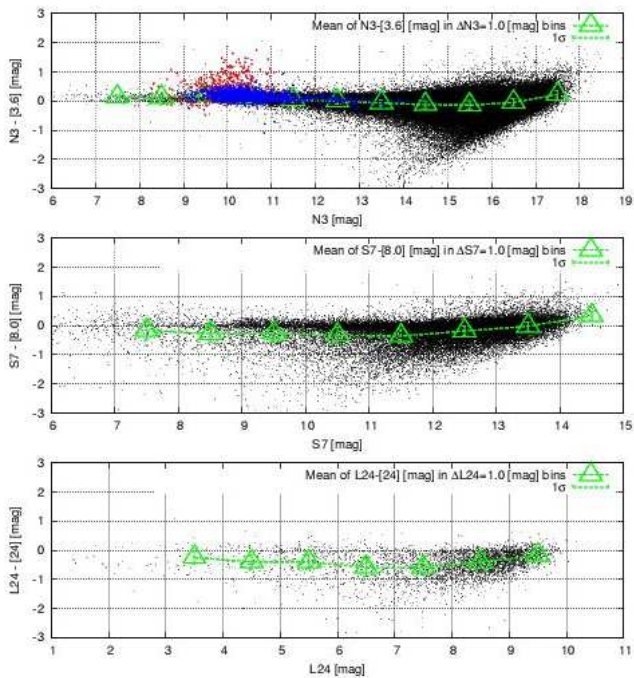}
  \end{center}
  \caption{Comparison between AKARI LMC survey and \textit{Spitzer} SAGE catalog using closely matched passbands (top panel: N3 vs [3.6], middle panel: S7 vs [8.0], bottom panel: L24 vs [24]). In the N3 vs [3.6] panel, the spectroscopically identified carbon stars (optical prism survey, \cite{kontizas2001}) are indicated by blue dots, and photometrically selected (see text for selection conditions) carbon-rich AGB candidates are shown in red dots.}
  \label{fig:hikakuSAGE}
\end{figure}

Although the bandpasses of the AKARI IRC and the \textit{Spitzer} IRAC/MIPS bands are different, comparison of closely matched bands is useful to test the calibration of the AKARI IRC data. We compared the IRC N3, S7, L24 photometries in our catalog with the corresponding 3.6, 8.0, and 24 $\mu$m fluxes of the sources in the \textit{Spitzer} SAGE catalog. Note that both of the IRC and IRAC absolute flux calibration assume a SED of $f_\nu \propto \nu^{-1}$, but MIPS flux scale assumes a source spectrum of a 10,000 K blackbody (\cite{mips2007}). Therefore we converted MIPS scale into IRC/IRAC one by adding $-$0.043 mag (\cite{bolatto2006}) to the MIPS 24 catalog magnitudes to make a direct comparison between L24 and MIPS 24 $\mu$m photometries. The distributions of the magnitude differences between the IRC and IRAC/MIPS fluxes as a function of the corresponding IRC magnitudes are shown in Figure~\ref{fig:hikakuSAGE}. The triangles show the mean of the residual magnitudes in the corresponding 1 mag bin, and the error bars show their 1 $\sigma$ standard deviations. Taking account of the differences in the bandpass, the AKARI IRC and IRAC/MIPS photometries appear consistent with each other. Quantitatively, they are in agreement within 9, 10, and 11 \% in N3 vs [3.6], S7 vs [8.0], and L24 vs [24] comparisons, respectively, using the sources with S/N $>$ 10 $\sigma$ and after subtracting systematic offsets.

In the comparison between N3 and [3.6] photometries, we find an interesting feature. It is seen that a group of bright sources (sources with $9 <$ N3 [mag] $< 11$) deviate from the overall trend. To explain the deviation, we cross-identified our sources with spectroscopically identified carbon stars (\cite{kontizas2001}), which are shown in blue dots. 
We photometrically select dusty carbon-rich AGB candidates that cannot be identified in the optical survey based on the following conditions.

\begin{itemize}
\item (J $-$ K$_s$) $>$ 2.0
\item K$_s$ $<$ (4/3)$\times$((J $-$ K$_s$) $-$ 2.0) + 12.0
\end{itemize}
These sources are indicated by red dots. The J and K$_s$ fluxes are obtained by cross-identifying the IRC catalog with the IRSF/SIRIUS LMC near-infrared catalog (\cite{kato2007}, see next section). It is true that the above conditions may include some dusty O-rich AGB stars, but such stars are small in number compared to the carbon-rich AGB stars. It is clearly seen that red dots are the sources that show the deviation in the diagram. The deviation should be attributed to the fact that the passband of the IRC N3 band is slightly bluer than that of the IRAC 3.6 $\mu$m, and it is more sensitive to the 3.1 $\mu$m HCN+C$_2$H$_2$ absorption feature, which is usually seen in carbon-rich AGB stars (e.g., \cite{yamamura2000}). The N3 data, thus, enable us to select sources with carbon-rich circumstellar chemistry on the basis of their colors.

\begin{table*}[htbp]
  \caption{Source Classification based on \cite{kraemer2002}.}\label{table:isosws}
  \begin{center}
    \begin{tabular}{llc}
    \hline
    \multicolumn{1}{c}{Grouped as} & \multicolumn{1}{c}{Classification in \cite{kraemer2002}} & Marks in Figure~\ref{fig:colcol2}. \\
    \hline
     Naked stars without salient molecular bands & 1.N & \textcolor{red}{\Large{+}} \\
     Naked stars with C-rich molecular bands & 1.NC & \textcolor{red}{\Large{$\times$}} \\
     Naked stars with O-rich molecular bands & 1.NO & \textcolor{red}{\Large{$\bullet$}} \\
     Naked stars with emission lines (Be stars) & 1.NE & \textcolor{red}{\Large{$\Box$}}\\
     Red giants with C-rich dust & 2.CE, 3.CE, 3.CR, 4.CR & \textcolor{green}{\Large{$\bullet$}} \\
     Red giants with O-rich dust & 2.SE, 3.SB, 3.SE & \textcolor{green}{\Large{$\times$}} \\
     Wolf-Rayet and R Corona Borealis stars & 3.W & \textcolor{blue}{\Large{+}} \\
     C-rich proto planetary nebulae (PPNe) & 4.CN, 4.CT & \textcolor{magenta}{\Large{$\bullet$}} \\
     Planetary nebulae (PNe) & 4.PN, 4.PU, 5.PN & \textcolor{cyan}{\Large{+}} \\
     OH/IR,PPNe,PNe,YSOs\footnotemark[1] & 4.SA, 4.SB, 4.SE, 4.SEC, & \Large{$\triangle$} \\
      & 4.U/SC, 5.E, 5.SA, 5.SE, 5.U \\
    \hline
    \end{tabular}
  \end{center}
  \footnotemark[1] Young stellar objects, such as T Tau stars and Herbig Ae/Be stars \\
\end{table*}

\section{General analysis using the AKARI LMC survey photometric catalog}
\subsection{Color-color diagram}
\subsubsection{Convolution of ISO SWS spectra}
With an aim to help an interpretation of what types of sources are detected by the AKARI survey, we take fully calibrated ISO SWS spectra data from the on-line database\footnote{http://isc.astro.cornell.edu/$^{\sim}$sloan/library/swsatlas/atlas.html} available (\cite{sloan2003}). Then we estimate their fluxes in the AKARI IRC system by convolving the ISO SWS spectra with the response curves of the IRC bands using the following equation,
\begin{equation}
f^\textrm{reference}_{\lambda_i} = \frac{\int \frac{R_i(\lambda)}{h\nu} f_\lambda(\lambda) d\lambda}{\int (\frac{\lambda_i}{\lambda})\frac{R_i(\lambda)}{h\nu}d\lambda} , 
\end{equation}
where the suffix $i$ denotes IRC bands, $\lambda_i$ represents the reference wavelengths of each IRC band tabulated in table~\ref{table:survey}, and $R_i(\lambda)$ indicates the spectral responses of each band. The calculated fluxes are further converted into the Vega-magnitudes. Note that the calculated magnitudes are in the IRC system (substituting $\lambda f_\lambda(\lambda) = \textrm{constant}$ to equation (1) gives $\lambda_i f^\textrm{reference}_{\lambda_i}$), which can be directly compared with the catalog values.

We divide ISO SWS sources into 10 main groups, based on the Table 6 in \citet{kraemer2002}. Sources that are classified as "uncertain classification" and "very uncertain classification" in \citet{kraemer2002} are excluded in the following discussions. Table~\ref{table:isosws} summarizes how we group the ISO SWS sources.

\begin{figure*}[htbp]
  \begin{center}
   \rotatebox{-90}{
   \FigureFile(120,120mm){figure8a.ps}
   }
   \FigureFile(170,170mm){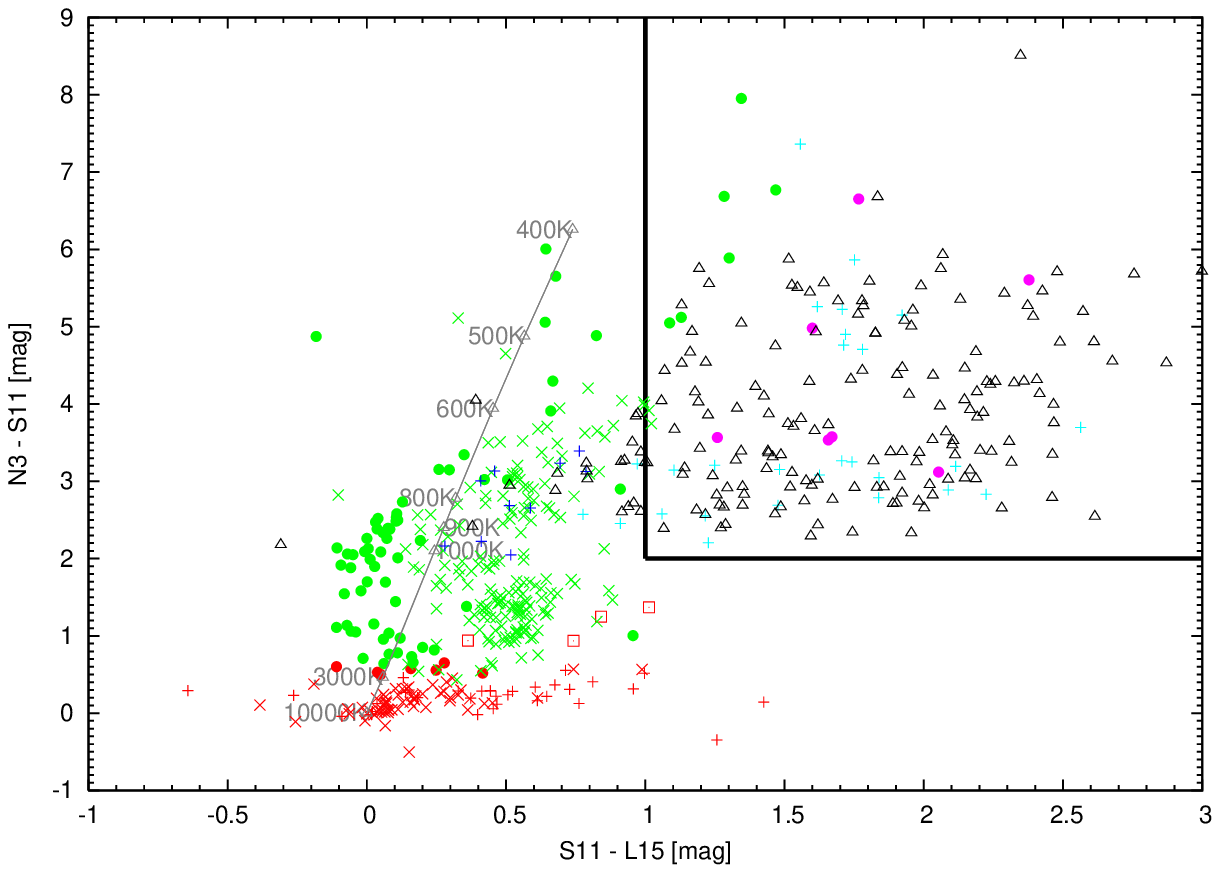}
  \end{center}
  \caption{The (S11 $-$ L15) vs (N3 $-$ S11) color-color diagram of AKARI/IRC sources (upper panel) and ISO SWS sources (lower panel). In the AKARI panel only sources with S/N $>$ 5 are included, and the (S11 $-$ L15, N3 $-$ S11) plane is binned by 0.05 $\times$ 0.05 mag$^2$ to calculate number of sources in each bin. The number density levels are logarithmic (see the wedge). The meanings of marks and their colors in ISO SWS panel are summarized in Table~\ref{table:isosws}. The solid gray line is a locus of the colors for blackbodies from temperature equals 400 to 10000K. The solid black lines show criteria for selecting YSO candidates (see section 6.2.2).}
  \label{fig:colcol2}
\end{figure*}

\subsubsection{(S11 $-$ L15) vs (N3 $-$ S11) diagram}
Figure~\ref{fig:colcol2} shows the color-color diagrams of AKARI sources (upper panel) and ISO SWS sources (lower panel) using colors of (S11 $-$ L15) and (N3 $-$ S11). In the AKARI diagram, only stars with S/N $>$ 5 in all employed three bands are included. The numbers of sources in each 0.05 $\times$ 0.05 mag$^2$ bin are counted and the fiducial color is applied according to their number densities (see the annotated color wedge). The solid gray line in the lower panel shows the blackbody locus with the temperature ranging from 400 to 10000K in the IRC system.

It is obvious that a wide variety of sources are detected in our survey. Comparing the two diagrams we can statistically separate various types of objects.
Two distinct main groups appear in the AKARI diagram: one centered around (S11 $-$ L15, N3 $-$ S11) of (0.1, 1.0) and the other centered around (1.0, 4.0). 

The former group includes two peaks, one centered around (0.2, 0.2) and the other centered around (0.25, 0.8). Based on a comparison with the ISO SWS classification, the sources in the first peak are stars without circmstellar dust, and the second peak is associated to red giants with circmstellar dust emission. 
ISO SWS spectra indicate that red giants with O-rich dust as well as C-rich dust may once become blue in the (S11 $-$ L15) color, and then turn red again in the course of their evolution. This can be explained by the change in strength of silicate or silicon carbide dust features in S11 band relative to the L15 flux.

According to the ISO SWS spectra, the red group may contain various types of objects such as PPNe, PNe, OH/IR stars, and YSOs. Also, the background galaxies may have similar colors (see the following), which makes the classification more difficult.
The NP spectroscopic data, and follow-up spectroscopic observations are needed to make clear classification of these objects.

\subsection{Color-magnitude diagrams}
An advantage to study sources in the LMC is that we can build color-magnitude diagrams (CMD) and estimate their absolute magnitudes by adding the distance modulus, which is reasonably well known (e.g., \cite{feast1987}). Based on the \textit{Spitzer} SAGE catalog, \citet{meixner2006} and \citet{blum2006} presented infrared color-magnitude diagrams of LMC sources. Here we add S11 and L15 data, which are unique to the AKARI survey. We use intriguing combinations of the IRC, IRAC\&MIPS and IRFS/SIRIUS bands and the results are shown in Figure~\ref{fig:colmag}. Units of vertical axes are the apparent magnitudes. It can be scaled to the absolute magnitude by subtracting the distance modulus. The corresponding wavebands for the vertical axes are indicated at the top of each panel. The employed colors for the horizontal axes are indicated at the bottom of each panel. All of the color-magnitude planes are binned by 0.1 $\times$ 0.1 mag$^2$ to calculate the number of sources in each bin, and the fiducial color is given according to the number density levels in a logarithmic scale (see the wedge at the bottom). Other than the trends pointed out in \citet{meixner2006} and \citet{blum2006}, new features are seen in our figure, primarily due to the addition of the S11 band data. 

\begin{figure*}[htbp]
  \begin{center}
   \rotatebox{-90}{
   \FigureFile(130,130mm){figure9.ps}
   }
  \end{center}
  \caption{Color magnitude diagrams of AKARI LMC sources using several combinations of IRC, IRAC, and IRSF/SIRIUS bands. The vertical axis is in apparent magnitude at the corresponding wavebands, which are indicated at the top of each panel. The allows indicate the newly found feature, which can be attributed to the red giants that have luminosities below the tip of the first red giant branch.}
  \label{fig:colmag}
\end{figure*}

\begin{figure*}[htbp]
  \begin{center}
   \rotatebox{-90}{
   \FigureFile(125,125mm){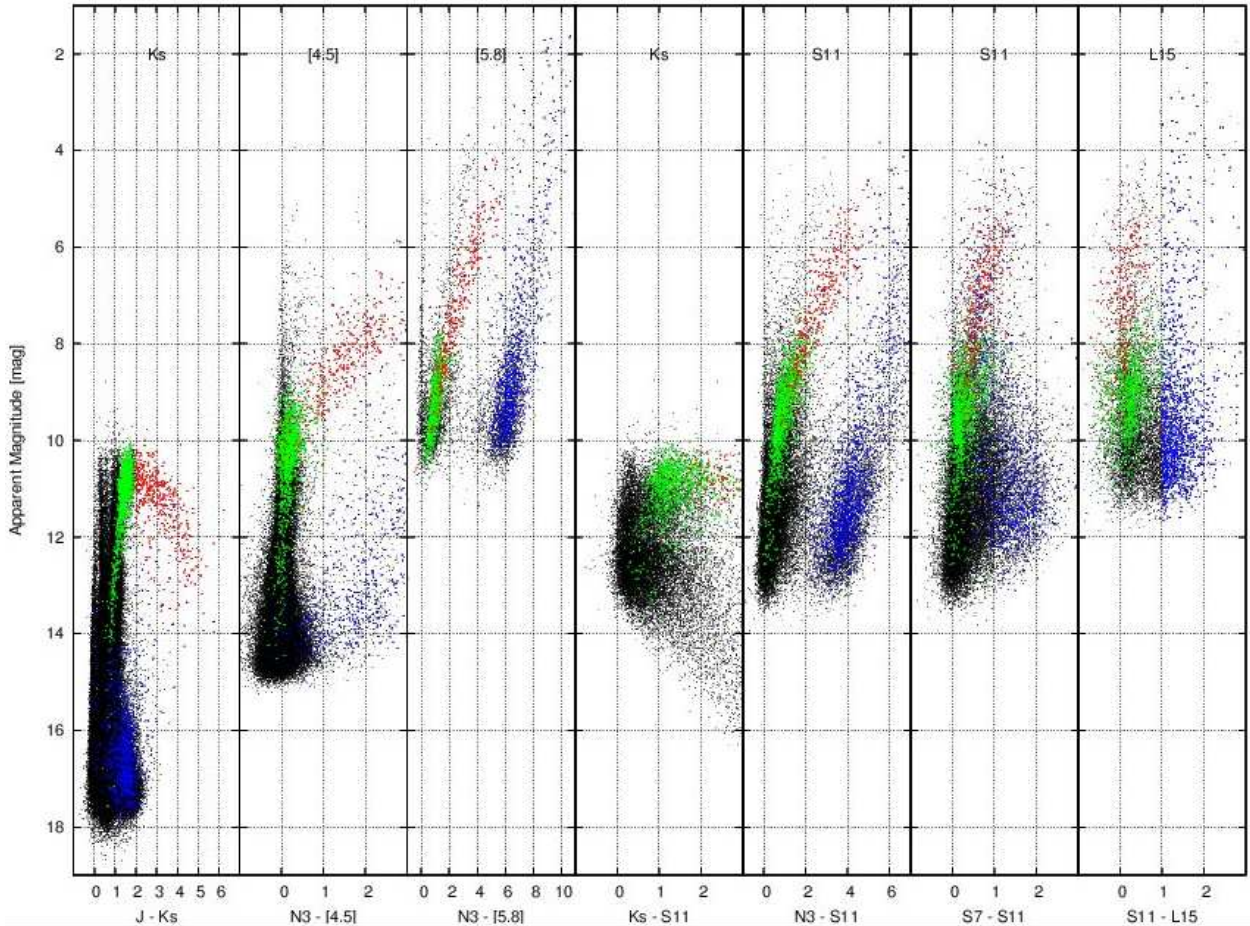}
   }
  \end{center}
  \caption{The same as Figure~\ref{fig:colmag}, but with spectroscopically confirmed carbon stars (green dots) by optical survey, candidates of dusty red giants (red dots), and candidates of YSOs (blue dots) are identified.}
  \label{fig:colmag2}
\end{figure*}

\begin{figure*}[htbp]
  \begin{center}
   \FigureFile(180,180mm){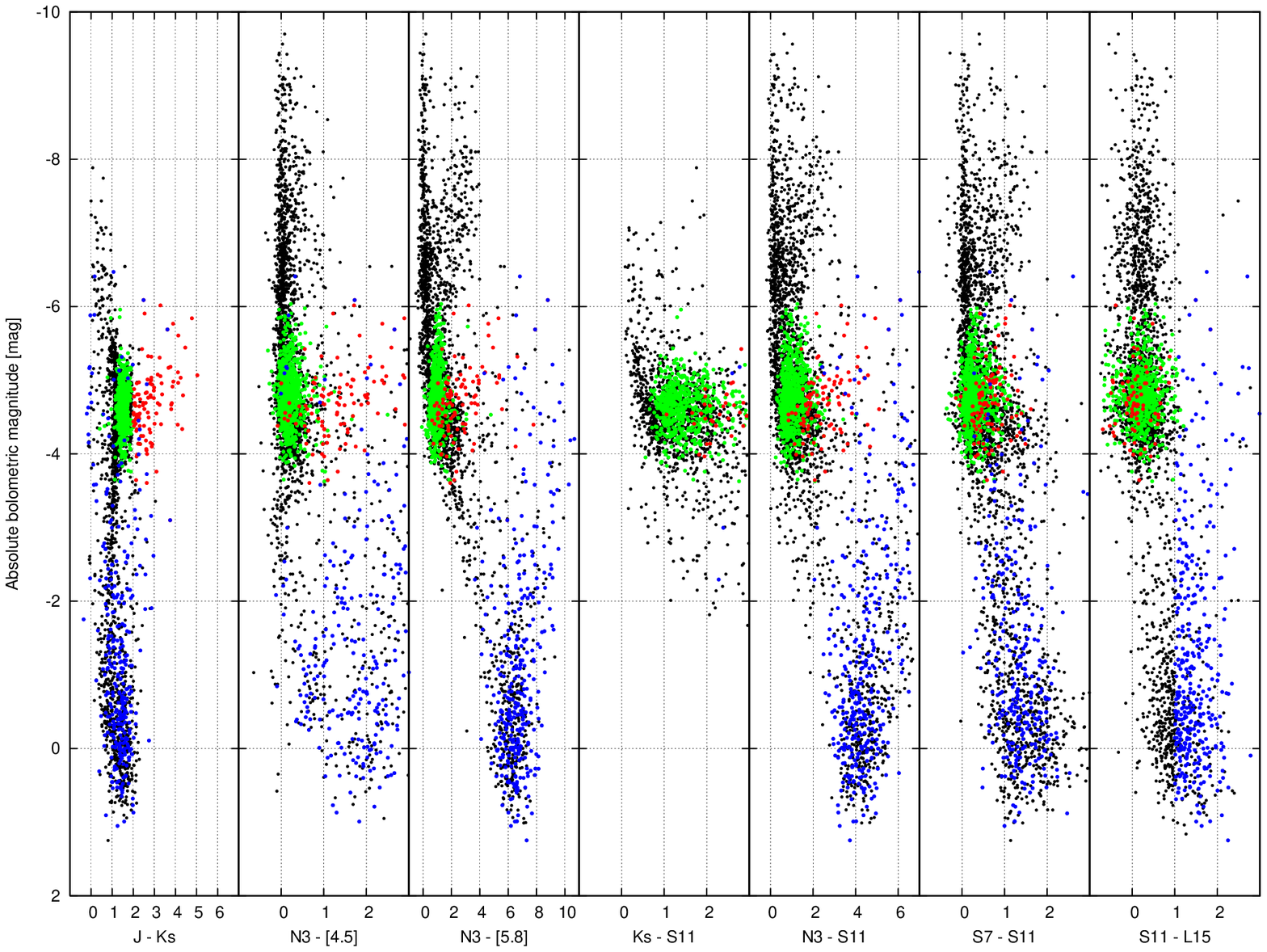}
  \end{center}
  \caption{The same as Figure~\ref{fig:colmag2} but with the vertical axis in absolute bolometric magnitude. We assumed a distance modulus of 18.5 mag to the Large Magellanic Cloud. The meanings of colors of marks are the same in Figure~\ref{fig:colmag2}.}
  \label{fig:colmag3}
\end{figure*}

\subsubsection{New sequence for faint red giants with circumstellar dust}
\citet{ramdani2001} showed that the ratio of the ISO 11.5 $\mu$m to DENIS K$_s$ 2 $\mu$m flux density is a good indicator of dust mass-loss from red giants. The (K$_s$ $-$ S11) vs K$_s$ panel indicates which red giants show circumstellar dust emission. It is clear that excess in S11 (K$_s$ $-$ S11 $>$ 0.5) is seen not only among the sources brighter than the tip of the first red giant branch (TRGB, K$_s$ $\sim$ 12.5mag, \cite{cioni2000b}), but also among the sources below the TRGB. Since the brighter sources exceed the K$_s$ luminosity of the TRGB, they should be an intermediate-age population and/or metal-rich old population AGB stars. On the other hand, the interpretation of the fainter sources is difficult. They can be metal-poor and old AGB stars that do not exceed the TRGB luminosity, or red giants on the first red giant branch (RGB star). It is well known that the number ratio of AGB to RGB stars near the TRGB should be about 1/3 for the intermediate age stars (\cite{renzini1992}). Thus, there are more RGB populations below the TRGB. The evidence for the mass-loss from the fainter sources is also seen in other panels. We see a bristle-shaped feature in (N3 $-$ S11) vs S11 panel, which is indicated by the arrow. The corresponding feature is marginal but also seen in the (S7 $-$ S11) panel (indicated by the arrow). 

\citet{lebzelter2006} obtained low-resolution mid-infrared (7.6 -- 21.7 $\mu$m) spectra of a star (V13) in the NGC104 with \textit{Spitzer}. The K$_s$ band luminosity of the star is fainter than the TRGB luminosity of NGC104. They showed that the star is devoid of the 9.7 $\mu$m emission band feature of amorphous silicate, but it has broad emission features at 11.5 $\mu$m (likely to be Al$_2$O$_3$) and 13 $\mu$m (likely to be an Al$-$O stretching vibration). Aluminium oxide features have been detected from low mass-loss rate oxygen-rich red giants (\cite{onaka1989, kozasa1997}). \citet{ita2007} suggested that the dust composition of V13 may be different from that of usual mass-losing AGB star (i.e., the brighter group). As it can be seen in Figure~\ref{fig:filters}, the IRC S11 band includes all of these emission features. Therefore, the feature indicated by the arrow in the (N3 - S11) vs S11 panel may be attributed to the red giants with aluminium oxide dust but without the silicate feature. 
\citet{blum2006} identified a group of red but faint O-rich red giants in their ([8.0] $-$ [24]) vs [8.0] color-magnitude diagram. The sources with S11 excess found in our (N3 $-$ S11) vs S11 panel may be similar to the red but faint O-rich red giants in \citet{blum2006}. Spectroscopic follow-up observations to identify the S11 excess with the Al$_2$O$_3$ band would be interesting.

\newcolumntype{d}[1]{D{.}{\cdot}{#1}}
\newcolumntype{.}{D{.}{.}{-1}}
\newcolumntype{,}{D{,}{,}{-1}}
\begin{table*}[htbp]
  \caption{Assumed zero magnitude fluxes and their corresponding reference wavelengths for the existing catalogs.}\label{table:zeromag}
  \begin{center}
    \begin{tabular}{c D..{2} D..{4} r}
    \hline
    Wavebands & \multicolumn{1}{c}{Zero magnitude flux} & \multicolumn{1}{c}{Reference wavelength} & Reference \\
              & \multicolumn{1}{r}{[Jy]} & \multicolumn{1}{r}{[$\mu$m]} &  \\
    \hline
    U & 1649 & 0.3745 & \citet{cohen2003a} \\
    B & 4060 & 0.4481 & " \\
    V & 3723 & 0.5423 & " \\
    I & 2459 & 0.8071 & " \\
    J & 1594 & 1.235  & \citet{cohen2003b} \\
    H & 1024 & 1.662  & " \\
    K$_s$   & 666.8 & 2.159 & "  \\
    $[3.6]$ & 280.9 & 3.550   & \citet{irac2006} \\
    $[4.5]$ & 179.7 & 4.493   & " \\
    $[5.8]$ & 115.0 & 5.731   & " \\
    $[8.0]$ & 64.1  & 7.872   & " \\
    $[24]$  & 7.14  & 23.68   & \citet{mips2007} \\
    \hline
    \end{tabular}
  \end{center}
\end{table*}

\subsubsection{Carbon stars, dusty red giants, and YSO candidates on the CMDs}
We identify carbon stars, dusty red giants, and YSO candidates by the following criteria, and estimate their location on the CMDs. The results are shown in Figures~\ref{fig:colmag2} and \ref{fig:colmag3}.
\begin{itemize}
    \item Optical carbon stars: As in section 5.4.3, we cross-identified our catalog with the carbon star catalog in the LMC (\cite{kontizas2001}) based on optical prism spectroscopy. They are shown by the green dots in Figures~\ref{fig:colmag2} and \ref{fig:colmag3}.
    \item Dusty red giants: We select sources that satisfy the same conditions employed in section 5.4.3. We use the term "dusty red giants" here, but a significant fraction of the sources in this category can be actually infrared carbon stars, which are elusive in the above optical spectroscopic survey. According to \citet{nikolaev2000}, stars with (J $-$ K$_s$) $>$ 2.0 in the LMC are primarily carbon-rich AGB stars. This can be explained as the J band flux of carbon-rich mass-losing red giants are attenuated by C$_2$ and CN absorptions \citep{loidl2001}. This criterion may be satisfied also by some dusty oxygen-rich AGB stars, but their number should be small compared to carbon stars. To confirm their circumstellar chemistry, our NP spectroscopic data would be effective. Sources in this category are shown by the red dots in Figures~\ref{fig:colmag2} and \ref{fig:colmag3}.
    \item YSO candidates: Based on the color-color diagram (Figure~\ref{fig:colcol2}), we select sources that satisfy (N3 $-$ S11) $>$ 2.0 and (S11 $-$ L15) $>$ 1.0. As seen in the ISO SWS panel of Figure~\ref{fig:colcol2}, sources selected by this criteria include not only YSOs but also other populations such as PPNe and PNe. Also, background galaxies may have colors similar to sources in this category (see the following). Sources in this category are shown by the blue dots in Figures~\ref{fig:colmag2} and \ref{fig:colmag3}.
\end{itemize}

Also, we calculated bolometric magnitudes for sources that satisfy the following conditions.
\begin{itemize}
\item Sources detected at least in two of four optical (U, B, V, and I) bands.
\item Sources detected at least in three of seven near-infrared (J, H, K$_s$, N3, [3.6], [4.5], and [5.8]) bands. 
\item Sources detected at least in three of four mid-infrared (S7, [8.0], S11, and L15) bands.
\item Sources detected at L24 and/or [24] bands.
\end{itemize}
Table~\ref{table:zeromag} summarizes the assumed zero magnitude fluxes and their corresponding reference wavelengths that we used to calculate the flux densities. For the IRC bands, we use the zero magnitude fluxes and reference wavelengths listed in Table~\ref{table:survey}. Since the zero magnitude fluxes for the IRSF/SIRIUS JHK$_s$ bands are not officially determined, we converted the IRSF system magnitudes into the 2MASS system by using the conversion equations provided in \citet{kato2007} before calculating the flux densities. Then we use a cubic spline to interpolate the spectral energy distributions and integrate them from the shortest wavelength at which the flux is available to 24 $\mu$m to obtain the apparent bolometric magnitudes. Finally, we scale them to the absolute magnitudes by assuming a distance modulus of 18.5 mag to the LMC. Figure~\ref{fig:colmag3} shows the result. For very red sources, such as galaxies, planetary nebulae, and extremely dusty AGB stars, the bolometric magnitudes may be underestimated to a large extent because the fluxes longward of 24 $\mu$m are not included. Therefore the bolometric magnitudes discussed here should be only lower limits. 

It is apparent that optical carbon stars, dusty red giants, and YSO candidates are well separated from each other by color and luminosity. Except for the near-infrared CMD, YSO candidates have redder colors than that of the others, and dusty red giants are brighter in the mid-infrared bands than optical carbon stars. In the (N3 $-$ [5.8]) vs [5.8] CMD, we see a group of sources that distribute along the vertical sequence around (N3 $-$ [5.8] $\sim$ 0.0). These sources are likely to be foreground Galactic stars that are classified as dwarf or giant. There is another feature in the diagram. We see a sequence for dusty red giants (red dots), which grows toward the redder and brighter part of the CMD. Above the sequence, we see another branch that grows toward the reddder and brighter part of the CMD. The brighter branch is clear in (N3 $-$ [5.8]) vs [5.8] diagram, and is also seen in the (N3 $-$ S11) vs S11 diagram. The corresponding bolometric magnitude CMDs tell us that most of dusty red giants have absolute bolometric magnitudes (M$_\textrm{bol}$) ranging from $-$4 to $-$5, while the sources on the brighter branch have M$_\textrm{bol}$ brighter than $-$6 mag and they are saturated in the IRSF/SIRIUS bands. It is suggested that sources on the brighter branch have stellar masses heavier than those of the sources on the sequence for dusty red giants. Their bolometric magnitudes agree well with the bolometric magnitudes for 88 red supergiants in the LMC (\cite{oestreicher1998}). Therefore we conclude that the sources on the brighter branch are red supergiants.

As we suggested in section 6.1.2, the (S11 $-$ L15) color of dusty red giants once becomes blue during their evolution due to the growth of silicate or silicon carbide emission in the S11 band. In fact most dusty red giants (red dots), if not all, are bluer in (S11 $-$ L15) than optical carbon stars (green dots).


\begin{table}[htbp]
  \caption{Galactic coordinates of reference fields.}\label{table:reference}
  \begin{center}
    \begin{tabular}{rrr}
    \hline
    ID & latitude & longitude \\
       & $[$degree$]$ & $[$degree$]$   \\
    \hline
    1 &  86.26824 &  34.38021 \\
    2 &  87.56487 &  34.71097 \\
    3 &  87.67876 &  35.02329 \\
    4 & 244.41766 & -34.71058 \\
    5 &  68.25446 &  35.19998 \\
    \hline
    \end{tabular}
  \end{center}
\end{table}

\begin{figure}[htbp]
  \begin{center}
   \FigureFile(85,85mm){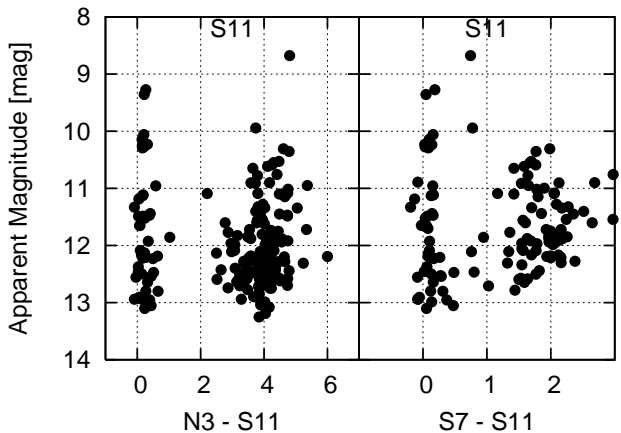}
  \end{center}
  \caption{Color-magnitude diagrams for sources in reference fields.}
  \label{fig:reference}
\end{figure}

\subsubsection{Background galaxies and foreground Milky Way sources}
To estimate how many background galaxies and foreground sources are expected in the CMDs, we use the leftover NIR and MIRS data that were kindly offered by the AKARI stellar working group (PI. Y.Nakada). The data were obtained when they took MIRL images of globular clusters. MIRL observations concurrently yield NIR and MIRS images that observe \timeform{25'} away from the MIRL center ($=$ target center). Also they observed some local group galaxies and took several reference images for them, which were about 1 degree away from the target centers. These reference data were also included in the following analysis. Among those offered by AKARI stellar working group, we select fields whose absolute value of galactic latitudes $|b|$ are in the range of 30$^\circ$ $<$ $|b|$ $<$ 40$^\circ$ that are comparable to the mean value of the AKARI LMC survey regions ($|b|$ $\sim$ 32.5$^\circ$). We reduce the data in the same way as we did for the LMC survey. The total area of the reference fields is 500 arcmin$^2$. The coordinates of the referenced fields are listed in Table~\ref{table:reference}. We check IRAS 24, 60 and 100 $\mu$m images of the corresponding coordinates, and found no appreciable indication of star forming regions and molecular clouds.

Using the data of the reference fields, we make CMDs that can be directly compared to Figures~\ref{fig:colmag} and \ref{fig:colmag2}. They are shown in Figure~\ref{fig:reference}. Most, if not all, of the sources in Figure~\ref{fig:reference} are expected to be background galaxies and foreground Milky Way stars, with a few possibility of stars in the outer skirts of target globular clusters. We check the tidal radii of the target globular clusters (\cite{harris1996}), and they are about a half of the offset value of \timeform{25'}. In the following analysis, we will ignore the small contribution from the stars in the outer skirts of each globular cluster. Due to the high source density, in particular in the central part of the LMC, the completeness limit of the LMC survey data must be shallower than that of the reference fields. Therefore, we only count the number of sources that satisfy the following conditions.

(1) For presumable background galaxy:
\begin{itemize}
\item (N3 $-$ S11) $>$ 2 [mag]
\item 10 $<$ S11 [mag] $<$ 12.2
\end{itemize}

(2) For presumable foreground stars:
\begin{itemize}
\item (N3 $-$ S11) $<$ 2 [mag]
\item 9 $<$ S11 [mag] $<$ 12.2
\end{itemize}

Then we found 72 background galaxies and 26 foreground stars in the reference fields. The differences in number counts for both populations among the reference fields are small. These results imply that we can expect about 500 background galaxies and 190 foreground stars in a square degree. This estimation agrees with the \textit{Spitzer} results (\cite{fazio2004b}, \cite{meixner2006}). On the other hand, we found 4483 and 14951 sources in our catalog (compiled from 10 deg$^2$ survey data) that satisfy the above conditions, respectively. Therefore most, if not all, of the sources that satisfy the condition (1) in our catalog should be background galaxies. Meanwhile, the fraction of foreground stars in the sources that satisfy condition (2) is estimated to be about 13\%.

\section{Summary}
We carried out an imaging and spectroscopic survey of a 10 deg$^2$ area of the LMC using IRC onboard AKARI. In this paper we discuss the imaging data obtained in the survey and describe a preliminary photometric catalog of bright point sources. The first point source catalog is planned to be released to the public in 2009. We cross-identified our catalog with the existing optical, near-infrared, and mid-infrared photometry catalogs, and used the combined data to build color-magnitude diagrams using several combinations of interesting wavebands and also to calculate absolute bolometric magnitudes. By virtue of the S11 and L15 bands that are unique to IRC, we found new interesting features in the color-magnitude diagrams. The present IRC data and the upcoming results from AKARI All-Sky survey together with existing radio and near-infrared ground-based observational results will provide a significant database to study the star-formation history and material circulation within a galaxy.

\section*{Acknowledgements}
We thank the referee for providing constructive comments and help in improving this paper. Y.I. thank Dr. Gregory C. Sloan for helpful comments at the IAU Symposium 256 held in the Keele University. We acknowledge the use of \textit{Spitzer} SAGE-LMC data in the assessment of the AKARI LMC results and thank the SAGE-LMC team, led by Dr. Margaret Meixner, for early access to data for this purpose. AKARI is a JAXA project with the participation of ESA. We are grateful to the members of AKARI stellar working group for providing us with their data, and also for valuable discussions that helped to improve this paper. This work is supported by a Grant-in-Aid for Scientific Research (A) No.~18204014 from Japan Society for the Promotion of Science and also supported by a Grant-in-Aid for Encouragement of Young Scientists (B) No.~17740120 from the Ministry of Education, Culture, Sports, Science and Technology of Japan. This work is partly supported by the JSPS grant (grant number 16204013). This publication makes use of data products from the Two Micron All Sky Survey, which is a joint project of the University of Massachusetts and the Infrared Processing and Analysis Center/California Institute of Technology, funded by the National Aeronautics and Space Administration and the National Science Foundation. This research has made use of the NASA/IPAC Infrared Science Archive, which is operated by the Jet Propulsion Laboratory, California Institute of Technology, under contract with the National Aeronautics and Space Administration.




\begin{thebibliography}{}
\bibitem[Aguirre et al.(2003)]{aguirre2003}
   Aguirre J.E., Bezaire J.J., Cheng E.S., et al., \ 2003, ApJ, 596, 273
\bibitem[Arendt et al.(1999)]{arendt1999}
   Arendt R.G., Dwek E., \& Moseley S.H. \ 1999, ApJ, 521, 234
\bibitem[ASTRO-F Observer's Manual.(2005)]{afobsman2005}
   ASTRO-F Observer's Manual, version 3.2, \ 2005 November 29, \\ http://www.ir.isas.jaxa.jp/ASTRO-F/Observation/ObsMan/afobsman32.pdf
\bibitem[Benjamin et al.(2003)]{benjamin2003}
   Benjamin R.A., Churchwell E., Babler B.L., et al., \ 2003, PASP, 115, 953
\bibitem[Bertin \& Arnouts(1996)]{bertin1996}
   Bertin E. \& Arnouts S. \ 1996, A\&AS, 117, 393
\bibitem[Blum et al.(2006)]{blum2006}
   Blum R.D., Mould J.R., Olsen K.A., et al., \ 2006, ApJ, 132, 2034
\bibitem[Bolatto et al.(2000)]{bolatto2000}
   Bolatto A.D., Jackson J.M., Israel F.P., et al., \ 2000, ApJ, 545, 234
\bibitem[Bolatto et al.(2006)]{bolatto2006}
   Bolatto A.D., Simon J.D., Stanimirovi\'{c} S., et al., \ 2006, ApJ, 655, 212
\bibitem[Borkowski et al.(2006)]{borkowski2006}
   Borkowski K.J., Williams B.J., Reynolds S.P., et al., \ 2006, ApJ, 642, 141
\bibitem[Chan \& Onaka.(2000)]{chan2000}
   Chan K.-W., \& Onaka T. \ 2000, ApJL, 533, L33
\bibitem[Cioni et al.(2000a)]{cioni2000a}
   Cioni M.-R., Loup C., Habing H.J., et al., \ 2000a, A\&AS, 144, 235
\bibitem[Cioni et al.(2000b)]{cioni2000b}
   Cioni M.R., van der Marel R.P., Loup C., et al., \ 2000b, A\&A, 359, 601
\bibitem[Clayton et al.(2001)]{clayton2001}
   Clayton D.D., Denault E.A.N., \& Meyer B.S., \ 2001, ApJ, 400, 222
\bibitem[Cohen et al.(2003a)]{cohen2003a}
   Cohen M., Megeath S., Hammersley T., et al, \ 2003a, AJ, 125, 2645
\bibitem[Cohen, Wheaton, \& Megeath(2003b)]{cohen2003b}
   Cohen M., Wheaton W.A., \& Megeath S.T., \ 2003b, AJ, 126, 1090
\bibitem[Cox \& Spaans(2006)]{cox2006}
   Cox N.L.J, \& Spaans M.,  \ 2006, A\&A, 451, 973
\bibitem[Crosas \& Menten(1997)]{crosas1997}
   Crosas M., \& Menten K., \ 1997, ApJ, 483, 913
\bibitem[Douvion et al.(2001)]{douvion2001}
   Douvion T., Lagage P.O., Cesarsky C.J., et al., \ 2001, A\&A, 373, 281
\bibitem[Dwek(1986)]{dwek1986}
   Dwek E. \ 1986, ApJ, 302, 363
\bibitem[Dwek(1998)]{dwek1998}
   Dwek E. \ 1998, ApJ, 501, 643
\bibitem[Egan et al.(2003)]{egan2003}
   Egan M.P., Price S.D., Kraemer K.E., et al., \ 2003, in Air Force Research Laboratory Technical Report, Vol. AFRL-VS-TR-2003-1589
\bibitem[Fazio et al.(2004a)]{fazio2004a}
   Fazio G.G., Hora J.L., Allen L.E., et al., \ 2004, ApJS, 154, 10
\bibitem[Fazio et al.(2004b)]{fazio2004b}
   Fazio G.G., Ashby M.L.N., Barmby P., et al., \ 2004, ApJS, 154, 39
\bibitem[Feast \& Walker(1987)]{feast1987}
   Feast M.W., \& Walker A.R. \ 1987, Ann. Rev. ARA\&A, 25, 345
\bibitem[Filipovic et al.(1998)]{filipovic1998}
   Filipovic M.D., Jones P.A., White G.L., et al., \ 1998, A\&AS, 130, 441
\bibitem[Fukui(2005)]{fukui2005}
   Fukui Y.,  \ 2005, IAUS, 227, 328
\bibitem[Gaustad et al.(2001)]{gaustad2001}
   Gaustad J.E., McCullough P.R., Rosing W., et al., \ 2001, PASP, 113, 1326
\bibitem[G\"{o}tz et al.(2006)]{gotz2006}
   G\"{o}tz D., Mereghetti S., Merlini D., et al., \ 2006, A\&A, 448, 873
\bibitem[Haberl \& Petsch(1999)]{haberl1999}
   Haberl F. \& Pietsch W. \ 1999, A\&AS, 139, 277
\bibitem[Haberl, Dennerl, \& Pietsch(2003)]{haberl2003}
   Haberl F., Dennerl K., \& Pietsch W., \ 2003, A\&A, 406, 471
\bibitem[Harris(1996)]{harris1996}
   Harris W.E., \ 1996, AJ, 112, 1487
\bibitem[Hosokawa \& Inutsuka(2006)]{hosokawa2006}
   Hosokawa T., \& Inutsuka S., \ 2006, ApJ, 646, 240
\bibitem[IRAC Data Handbook(2006)]{irac2006}
   IRAC Data Handbook, version 3.0, \ 2006 January 20, \\ http://ssc.spitzer.caltech.edu/mips/dh \\ /mipsdatahandbook3.3.pdf
\bibitem[Ishihara et al.(2006)]{ishihara2006}
   Ishihara D., Wada T., Onaka T., et al., \ 2006, PASP, 118, 324
\bibitem[Israel \& Schwering(1986)]{israel1986}
   Israel F.P., \& Schwering P.B., et al., \ 1986, ASSL, 124, 383
\bibitem[Ita et al.(2004a)]{ita2004a}
   Ita Y., Tanab\'{e} T., Matsunaga N., et al., \ 2004a, MNRAS, 347, 720
\bibitem[Ita et al.(2004b)]{ita2004b}
   Ita Y., Tanab\'{e} T., Matsunaga N., et al., \ 2004b, MNRAS, 353, 705
\bibitem[Ita et al.(2007)]{ita2007}
   Ita Y., Tanab\'{e} T., Matsunaga N., et al., \ 2007, PASJ, 59, 437
\bibitem[Jones et al.(1994)]{jones1994}
   Jones A.P., Tielens A.G.G.M., Hollenbach D.J., et al., \ 1994, ApJ, 433, 797
\bibitem[Jones et al.(1996)]{jones1996}
   Jones A.P., Tielens A.G.G.M., Hollenbach D.J., et al., \ 1996, ApJ, 469, 740
\bibitem[Kato et al.(2007)]{kato2007}
   Kato D., Nagashima C., Nagayama T., et al., \ 2007, PASJ, 59, 615
\bibitem[Kawada et al.(2007)]{kawada2007}
   Kawada M., Baba H., Barthel P.D., et al., \ 2007, PASJ, 59, S389
\bibitem[Kennicutt \& Hodge(1986)]{kennicut1986}
   Kennicutt R.C., \& Hodge P.W., \ 1986, ApJ, 306, 130
\bibitem[Kepler et al.(2007)]{kepler2007}
   Kepler S.O., Kleinman S.J., Nitta A., et al., \ 2007, MNRAS, 375, 1315
\bibitem[Kim et al.(1998)]{kim1998}
   Kim S., Staveley-Smith L., Dopita M.A., et al., \ 1998, ApJ, 503, 674
\bibitem[Kontizas et al.(2001)]{kontizas2001}
   Kontizas E., Dapergolas A., Morgan D.H., Kontizas M., \ 2001, A\&A, 369, 932
\bibitem[Koo et al.(2007)]{koo2007}
   Koo B.-C., Lee H.-G., Moon D.-S., et al., \ 2007, PASJ, 59, S455
\bibitem[Koo et al.(2008)]{koo2008}
   Koo B.-C., McKee C.F., Lee J.-J., et al., \ 2008, ApJ, 673, 147
\bibitem[Koornneef(1982)]{koornneef1982}
   Koornneef J., \ 1982, A\&A, 107, 247
\bibitem[Kozasa \& Sogawa(1997)]{kozasa1997}
   Kozasa T., Sogawa H., \ 1997, Ap\&SS, 251, 165
\bibitem[Kraemer et al.(2002)]{kraemer2002}
   Kraemer K.E., Sloan G.C., Price S.D., et al., \ 2002, ApJS, 140, 389
\bibitem[Krause et al.(2004)]{krause2004}
   Krause O., Birkmann S.M., Rieke G.H., et al., \ 2004, Nature, 432, 596
\bibitem[Kumai, Basu, \& Fujimoto(1993)]{kumai1993}
   Kumai Y., Basu B., \& Fujimoto M., \ 1993, ApJ, 404, 144
\bibitem[Lebzelter et al.(2006)]{lebzelter2006}
   Lebzelter T., Posch T., Hinkle K., et al., \ 2006, ApJ, 653, L145
\bibitem[Loidl et al.(2001)]{loidl2001}
   Loidl R., Lancon A., Jorgensen U. G., \ 2001, A\&A, 371, 1065
\bibitem[Long, Helfand, \& Grabelsky(1981)]{long1981}
   Long K.S., Helfand D.J., \& Grabelsky D.A., \ 1981, ApJ, 248, 925
\bibitem[Lorente et al.(2007)]{lorente2007}
   Lorente R., Onaka T., Ita Y., et al., \ 2007, AKARI IRC Data User's Manual ver. 1.3, http://www.ir.isas.jaxa.jp/AKARI/Observation/
\bibitem[Luks \& Rohlfs(1992)]{luks1992}
   Luks Th., \& Rohlfs K., \ 1992, A\&A, 263, 41L
\bibitem[Matsuhara et al.(2007)]{matsuhara2007}
   Matsuhara H., Wada T., Pearson C.P., et al., \ 2007, PASJ, 59, S543
\bibitem[Meikle et al.(2007)]{meikle2007}
   Meikle W.P.S., Mattila S., Pastorello A., et al., \ 2007, ApJ, 665, 608
\bibitem[Meixner et al.(2006)]{meixner2006}
   Meixner M., Gordon K.D., Indebetouw R., et al., \ 2006, AJ, 132, 2288
\bibitem[MIPS Data Handbook(2007)]{mips2007}
   MIPS Data Handbook, version 3.3, \ 2007 August 07, \\ http://ssc.spitzer.caltech.edu/mips/dh \\ /mipsdatahandbook3.3.pdf
\bibitem[Mizuno et al.(2001)]{mizuno2001}
   Mizuno N., Yamaguchi R., Mizuno A., et al., \ 2001, PASJ, 53, 971
\bibitem[Mochizuki et al.(1994)]{mochizuki}
   Mochizuki K., Nakagawa T., Doi Y., et al., \ 1994, ApJ, 430, 37
\bibitem[Murakami et al.(2007)]{murakami2007}
   Murakami H., Baba H.,  Barthel P., et al., \ 2007, PASJ, 59, S369
\bibitem[Nikolaev \& Weinberg(2000)]{nikolaev2000}
   Nikolaev S., \& Weinberg M.D., \ 2000, ApJ, 542 804
\bibitem[Oestreicher \& Schmidt-Kaler(1998)]{oestreicher1998}
   Oestreicher M.O., \& Schmidt-Kaler Th., \ 1998, MNRAS, 299, 625
\bibitem[Ohyama et al.(2007)]{ohyama2007}
   Ohyama Y., Onaka T., Matsuhara H., et al., \ 2007, PASJ, 59, S411
\bibitem[Onaka et al.(1989)]{onaka1989} 
   Onaka T., De Jong T., Willems F.J., \ 1989, A\&A, 218, 1690
\bibitem[Onaka(2000)]{onaka2000}
   Onaka T. \ 2000, Adv. Sp. Res. 25, 2167
\bibitem[Onaka et al.(2007)]{onaka2007} 
   Onaka T., Matsuhara H., Wada T., et al., \ 2007, PASJ, 59, S401
\bibitem[Onaka et al.(2008)]{onaka2008} 
   Onaka T., Roellig T.L., Okada Y., et al., \ 2008, ASP Conf. ser. 381, 80
\bibitem[Perryman et al.(1997)]{perryman1997} 
   Perryman M.A.C., Lindegren L., Kovalevsky J., et al., \ 1997, A\&A, 323, L49  
\bibitem[Ramdani \& Jorissen(2001)]{ramdani2001}
   Ramdani A., \& Jorissen A. \ 2001, A\&AS 372, 85
\bibitem[Renzini(1992)]{renzini1992}
   Renzini A., \ 1992, in Barbuy B., \& Renzini A., eds, ``The Stellar Populations of Galaxies'', Kluwer, Dordrecht, p.325   
\bibitem[Rho et al.(2008)]{rho2008}
   Rho J., Kozasa T., Reach W.T., et al., \ 2008, ApJ, 673, 271
\bibitem[Rieke et al.(2004)]{rieke2004}
   Rieke G.H., Young E.T., Engelbracht C.W., et al., \ 2004, ApJS, 154, 25
\bibitem[Sakon et al.(2004)]{sakon2004}
   Sakon I., Onaka T., Ishihara D., et al., \ 2004, ApJ, 609, 203
\bibitem[Sakon et al.(2005)]{sakon2005}
   Sakon I., Onaka T., Kaneda H., et al., \ 2005, IAUS, 235, 145
\bibitem[Sasaki et al.(2000)]{sasaki2000}
   Sasaki M., Haberl F., Pietsch W., et al., \ 2000, A\&AS, 143, 391
\bibitem[Seok et al.(2008)]{seok2008}
   Seok J.Y., Koo B.-C., Onaka T., et al., \ 2008, PASJ, submitted
\bibitem[Skrutskie et al.(2006)]{skrutskie2006} 
   Skrutskie M.F., Cutri R.M., Stiening R., et al., \ 2006, AJ, 131, 1163
\bibitem[Sloan et al.(2003)]{sloan2003} 
   Sloan G.C., Kraemer K.E., Price S.D., et al., \ 2003, ApJS, 147, 379
\bibitem[Smith, Cornett, \& Hill(1987)]{smith1987} 
   Smith A.M., Cornett R.H., \& Hill R.S., \ 1987, ApJ, 320, 609
\bibitem[Tappe, Rho, \& Reach(2006)]{tappe2006}
   Tappe A., Rho J., and Reach W.T., \ 2006, ApJ, 653, 267
\bibitem[Temim et al.(2006)]{temim2006}
   Temim T., Gehrz R.D., Woodward C.E., et al., \ 2006, AJ, 132, 1610 
\bibitem[Todini \& Ferrara(2001)]{todini2001}
   Todini P. \& Ferrara A., \ 2001, MNRAS, 325, 726
\bibitem[Udalski, Kubiak \& Szymanski(1997)]{udalski1997}
   Udalski A., Kubiak M., Szyma\'{n}ski M., \ 1997, AcA, 47, 319
\bibitem[van Loon et al.(2005)]{vanloon2005}
   van Loon J.Th., Oliveira J.M., Wood P.R., et al., \ 2005, MNRAS, 364, 71
\bibitem[Werner et al.(2004)]{werner2004} 
   Werner M.W., Roellig T.L., Low F.J., et al., \ 2004, ApJS, 154, 1
\bibitem[B. Williams et al.(2006)]{bwilliams2006} 
   Williams B.J., Borkowski K.J., Reynolds S.P., et al., \ 2006, ApJ, 652, 33
\bibitem[R. Williams, Chu, \& Gruendl(2006)]{rwilliams2006} 
   Williams R.M., Chu Y.-H., and Gruendl R. \ 2006, AJ, 132, 1877
\bibitem[Yamaguchi et al.(2001)]{yamaguchi2001} 
   Yamaguchi R., Mizuno N., Onishi T., et al., \ 2001, PASJ, 53, 959
\bibitem[Yamamura \& de Jong(2000)]{yamamura2000} 
   Yamamura I., \& de Jong T. \ 2000, ESASP, 456, 155
\bibitem[Zaritsky et al.(2004)]{zaritsky2004}
   Zaritsky D., Harris J., Thompson I.B., et al., \ 2004, AJ, 128, 1606
\bibitem[\.{Z}ebru\'{n} et al.(2001)]{zebrun2001}
   \.{Z}ebru\'{n} K., Soszynski I., Wozniak P.R., et al., \ 2001, AcA, 51, 317




\end{thebibliography}
\end{document}